\documentclass[
    amsmath,
    amssymb,
    aps,
    prx,
    twocolumn,
    10pt,
    floatfix,
    superscriptaddress
]{revtex4-2}

\usepackage{customarticlestyle}
\usepackage{soul}

\newcommand{\hexafoil}[0]{\rm hfoil}
\newcommand{\heavyhex}[0]{\rm hhex}

\begin{document}
\title{\gettitle}
\date{\today}

\author{Anthony Gandon$^{\orcidanthony}$}
\email{anthony.gandon@ibm.com}
\affiliation{\ibmquantum}
\affiliation{\ethzurich}
\author{Alessandro Mariani$^{\orcidalessandro}$}
\affiliation{\torino}
\author{Debasish Banerjee$^{\orciddebasish}$}
\affiliation{\southampton}
\author{Emilie Huffman$^{\orcidemilie}$}
\affiliation{\wakeforest}
\author{Gurtej Kanwar$^{\orcidtej}$}
\affiliation{\univedinburgh}
\author{Francesco Tacchino$^{\orcidfrancesco}$}
\affiliation{\ibmquantum}
\author{Uwe-Jens Wiese$^{\orciduwe}$}
\affiliation{\bern}
\author{Ivano Tavernelli$^{\orcidivano}$}
\affiliation{\ibmquantum}

\begin{abstract}
    The (2+1)D $\mathrm{U}(1)$ pure gauge theory always exists in the confining phase, with strings of non-zero string tension giving a characteristic linear potential between static charges. This makes it a useful testing ground for quantum computing methods designed to study string dynamics of confining gauge theories.
    Here we implement a minimal \uone~quantum link model on a quantum computer with qubit degrees of freedom representing the dual height variables of the model.
    A specifically tailored lattice geometry is chosen to match the heavy-hexagonal geometry of the IBM quantum hardware used here, minimizing non-adjacent qubit interactions.
    This facilitates an efficient realization of plaquette interactions and enables effective calculations of real-time dynamics that are inaccessible to traditional quantum Monte Carlo.
    By performing quantum quenches from a simple initial string state, we probe the transverse quantum fluctuations of the string before it thermalizes. 
    Our experimental results from digital quantum simulations, with up to 112 qubits, show good agreement with reference tensor-network calculations at short times and with thermal averages at long times. Near the phase transition, the dynamics exhibit large fluctuations of the initial string that extend across both spatial dimensions of the lattice.  
    Nonetheless, our error-mitigated estimators from the quantum hardware also give accurate predictions in that regime, with noise-induced violations of local gauge symmetries comparable to finite-bond-dimension tensor-network results.
\end{abstract}

\maketitle

\section{Introduction} \label{sec:intro}
The standard model of particle physics comprises everything that we know today about the fundamental constituents of matter and their electroweak and strong interactions. Most physical processes of the electroweak interaction can be addressed in the framework of perturbation theory using Feynman diagrams. On the other hand, the strong force is given by Quantum Chromodynamics (QCD), the $\mathrm{SU}(3)$ gauge theory describing interactions between quarks mediated by the gluon field. Being strongly coupled at low energies, its study requires the use of non-perturbative methods. The only known method that allows us to treat QCD from first principles is lattice field theory~\citep{DeGrand2006LatticeMethodsQuantum}. 
By mapping $d$-dimensional quantum Hamiltonians onto $(d+1)$-dimensional classical statistical mechanical systems, the physical properties of QCD can be studied using powerful Monte Carlo techniques at both zero and finite temperatures.
Remarkably accurate results have been obtained in characterizing the static properties of hadrons (i.e., baryons and mesons), the behavior of the deconfined quark--gluon plasma at high temperatures, and the nature of the associated transition~\citep{Aoki2026FLAGReview2024,Borsanyi2023EquationStateHotanddense,Aarts2023PhaseTransitionsParticle}.
Despite all these successes, there remain interesting and important open questions that cannot be addressed with Monte Carlo. These include the real-time evolution of systems out of thermal equilibrium, such as the expansion of a droplet of quark-gluon plasma that emerges from an ultra-relativistic heavy-ion collision~\citep{Banuls2020SimulatingLatticeGauged,DiMeglio2024QuantumComputingHighenergy,Halimeh2025QuantumSimulationOutofequilibrium}. In such cases, sign or complex action problems prevent
controlled importance-sampling estimates in polynomial time~\citep{Troyer2005ComputationalComplexityFundamentala}, which underlie Monte Carlo methods. 

Recent developments in engineering quantum simulators, i.e.\ quantum many-body platforms with controllable interactions, have opened new paths for understanding the dynamics of quantum systems out of equilibrium~\citep{Martinez2016RealtimeDynamicsLatticeb,Bernien2017ProbingManybodyDynamicsa,Ercolessi2018PhaseTransitionsZ,Ebadi2021QuantumPhasesMatter,Surace2020LatticeGaugeTheoriesa,Yang2020ObservationGaugeInvariance,Zhou2022ThermalizationDynamicsGaugea,Semeghini2021ProbingTopologicalSpin,Moss2024EnhancingVariationalMonte,Schuhmacher2025ObservationHadronScatteringa,Cobos2025RealtimeDynamics2+1Da,Gonzalez-Cuadra2025ObservationStringBreaking,Mariani2023HamiltoniansGaugeinvariantHilbert,Luo2025QuantumSimulationBubble,Davoudi2025QuantumComputationHadron,Klco2020SU2NonAbelianGaugea,Maiti2025SpontaneousSymmetryBreaking,Farrell2024QuantumSimulationsHadron,Farrell2024QuantumSimulationsHadron,Huffman2022RealtimeEvolutionGaugeinvariant,Ilcic2026ObservationRobustCoherent,Mou2025ScalableQuantumComputation}. 
Multiple approaches have been proposed for simulating lattice gauge theories (LGTs) on quantum computers~\citep{Wiese2013UltracoldQuantumGases,Chandrasekharan1997QuantumLinkModelsa,Brower1999QCDQuantumLink,Zohar2015QuantumSimulationsLattice,Zache2022ContinuumLimitD1,Ciavarella2024QuantumSimulationSU3a,Raychowdhury2020LoopStringHadron,Kadam2023LoopstringhadronFormulationSU3,Banerjee2013AtomicQuantumSimulationa,Fontana2025EfficientFiniteresourceFormulation,Jakobs2025DynamicsHamiltonianLattice,Zache2023QuantumClassicalSpinnetwork,Alexandru2024FuzzyGaugeTheory,Gustafson2024PrimitiveQuantumGates,Liu2022QubitRegularizationQubit,Bergner2024QCDQuantumComputer,Ciavarella2025TruncationUncertaintiesAccurate}, via either analog schemes or digital quantum computation.

In this work, we focus on the simulation of lattice gauge theories on digital quantum computers based on superconducting qubits. Given the fixed connectivity and limited coherence times of current hardware, we restrict our attention to theories in two spatial dimensions with a reduced number of field degrees of freedom. Specifically, we consider the pure gauge sector of an Abelian $\mathrm{U}(1)$ LGT within the quantum link model (QLM) framework~\citep{Wiese2022QuantumLinkModelsa,Chandrasekharan1997QuantumLinkModelsa}, where exact gauge invariance is realized in a finite-dimensional local Hilbert space compatible with spin degrees of freedom. QLMs have been successfully employed to regularize QCD itself~\citep{Wiese2022QuantumLinkModelsa}, and they offer a scalable route to constructing gauge-invariant theories for continuous groups without requiring large bosonic resources~\citep{Zache2022ContinuumLimitD1}. Notably, the minimal gauge-invariant formulation involves only a single qubit per link, in contrast to the Wilson formulation~\citep{Wilson:1974sk,Kogut1975HamiltonianFormulationWilsonsa}. 
We further perform an exact duality transformation in which the gauge fields are mapped onto height variables defined at the centers of spatial plaquettes~\citep{Banerjee201321dU1,Celi2020EmergingTwoDimensionalGaugeb}. This dual representation significantly simplifies the implementation of plaquette interactions, which are otherwise a major challenge in quantum simulations~\citep{Weimer2010RydbergQuantumSimulator,Zohar2012SimulatingCompactQuantum,Mezzacapo2015NonabelianSU2Lattice,Dai2017FourbodyRingexchangeInteractions}. In the dual picture, the plaquette operator reduces to a single conditional flip of a height variable, whereas in the electric-flux basis it corresponds to an off-diagonal term involving a coherent transformation of all fluxes around a plaquette.

QLMs on square and triangular lattices have been previously studied in equilibrium with highly efficient cluster algorithms~\citep{Banerjee201321dU1,Banerjee2022NematicConfinedPhasesa}, as well as large scale exact diagonalisation (ED) and DMRG methods~\citep{Shannon2004CyclicExchangeIsolated,Tschirsich2019PhaseDiagramConformal,Stornati2023CrystallinePhasesFinite}. 
On both lattices, there exist distinct confined phases that are characterized by spontaneously broken charge conjugation or lattice translation and rotation symmetries. 
Remarkably, the interfaces, which separate distinct confined phases, manifest themselves as unbreakable confining strings that can only end in external static charges. A generic feature of these models is flux fractionalization: the electric flux string connecting two external charges $\pm 1$ can split into two strands, each carrying a fractionalized flux of $1/2$. Since their string dynamics are even more multi-faceted than those of conventional Wilson-type lattice gauge theories, QLMs are ideally suited for exploring the capabilities of quantum hardware to simulate the behavior of the corresponding strings in real time. 
Our experiments on quantum many-body dynamics leverage IBM quantum processors featuring a heavy-hexagonal qubit topology~\citep{Kim2023EvidenceUtilityQuantumb}.
To avoid unnecessary digital routing of qubit interactions, we design a QLM pure gauge theory on a tailor-made lattice geometry, defined such that the dual Hamiltonian, acting on plaquette height variables, is composed of three- and four-body qubit operators that are nearest-neighbors on the heavy-hexagonal connectivity graph of an IBM Heron device. 
The resulting pure gauge theory lives on a hexafoil lattice, that, while unconventional, is well-motivated. As depicted in~\Cref{fig:hexafoillattice_dualization}, the faces of this lattice consist of \emph{triangle} plaquettes, bounded by three links, adjacent to \emph{petal} plaquettes, which are bounded by two links. 
Notably, triangle and petal plaquettes are not related to each other by lattice translation or rotation. We study the nature of the confining phases in this model, as well as the properties of flux strings joining static charges.

The paper is organized as follows. In~\Cref{sec:model}, we introduce the model on the hexafoil geometry and describe the duality mapping that enables a hardware-efficient encoding on the heavy-hex qubit architecture. In~\Cref{sec:quench}, we present our experimental setup and results for digitized quantum quenches starting from product-state string configurations and spanning multiple phases of the model. These experiments allow us to probe the prethermalization dynamics of strings in quantum link gauge theories, a regime inaccessible to classical Monte Carlo methods. At the phase transition point, we observe a broadening of the string’s transverse fluctuations and discuss the classical simulability of the phenomena. Finally, in~\Cref{sec:static_phases}, we relate the string oscillations observed in the dynamics with the properties of the confining string in equilibrium using numerical QMC simulations.

\section{Model and setup} \label{sec:model}

\subsection{Spin-\texorpdfstring{$1/2$}{1/2} \texorpdfstring{\uone}{U(1)} QLM on the hexafoil lattice}
 The Hamiltonian for a pure gauge \uone~LGT is given by
 \begin{align}
     \mathcal{H} = \frac{g^2}{2} \sum_{x,\hat{i}} E^2_{x,\hat{i}} 
     + \frac{1}{4g^2} \sum_{\Box} (U_\Box + U^\dagger_\Box),
\end{align}
where $E_{x,\hat{i}}$ is the electric flux operator associated to the link between $x$ and $x+\hat{i}$ and $U_\Box$ is the gauge-invariant plaquette term on the plaquette labeled by $\Box$.
The Kogut-Susskind formulation~\citep{Kogut1975HamiltonianFormulationWilsonsa} uses quantum rotors as degrees of freedom, whereby on each link $(x, \hat{i})$ one has an element $U_{x, \hat{i}} \in U(1)$. The products of $U_{x, \hat{i}}$ on the oriented links at the boundary of a plaquette $\Box$ then form the operator $U_\Box$. The infinite-dimensional Hilbert space of that theory on each link is labelled by the integer-valued eigenvalues of the operator $E_{x,\hat{i}}$, which satisfies canonical commutation relations
\begin{equation}\label{eq:commutation relations}
    [E_{x, \hat{i}}, U_{y, \hat{j}}] = U_{x, \hat{i}} \delta_{xy} \delta_{ij} \ .
\end{equation}
Most implementations of this theory, numerical or experimental, involve a truncation either in the electric field or in the magnetic field basis.
This must necessarily in some way violate the commutation relations~\labelcref{eq:commutation relations}, which cannot be realized in a finite-dimensional Hilbert space. 
In this work we instead use the quantum link formulation~\citep{Chandrasekharan1997QuantumLinkModelsa, Wiese2022QuantumLinkModelsa}, where, on each link, one replaces $E,U,U^\dagger$ with spin operators, $E\to S^3, U\to S^+, U^\dagger \to S^-$. These satisfy the commutation relation~\Cref{eq:commutation relations} exactly, and therefore the resulting theory has an exact gauge symmetry, at the expense of the unitarity of $U$. The three operators $S^3, S^\pm$ form an $\mathfrak{su}(2)$ algebra on each link and admit finite-dimensional representations of arbitrary spin~$s$. Here we choose the minimal representation $s=1/2$, so the spin operators can be represented via the Pauli matrices $X, Y, Z$.
Despite the $\mathfrak{su}(2)$ embedding algebra, the gauge symmetry remains {\uone}.
 
\begin{figure}[t]
    \centering
    \includegraphics[width=0.9\linewidth]{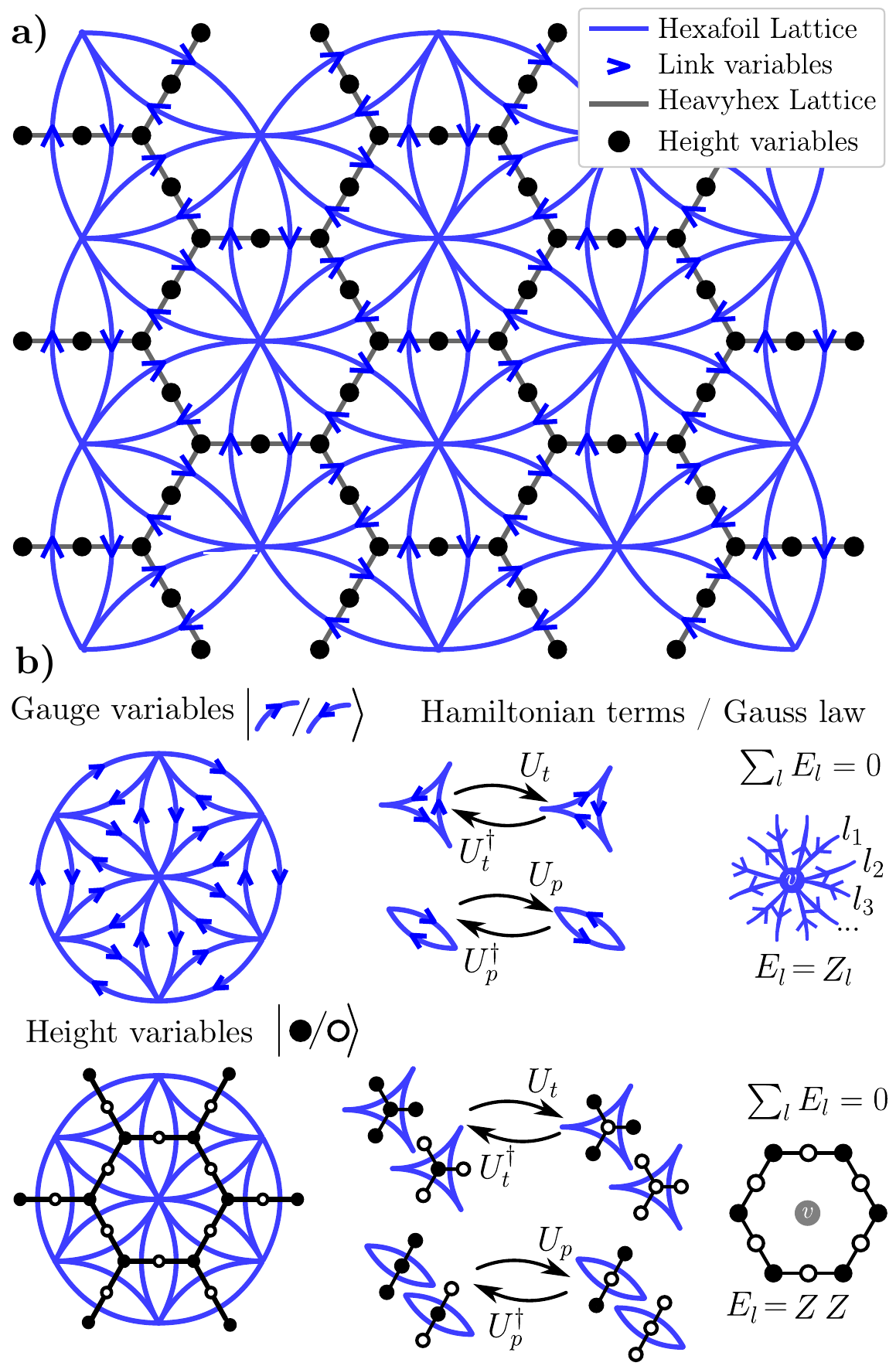}
    \caption{a) Hexafoil lattice, in blue, with gauge degrees of freedom on the edges of the hexafoil lattice. The dual heavy-hexagonal lattice, in black, corresponds to the available hardware topology. It has qubit degrees of freedom on the vertices represented by dots. b) Correspondence of states and operators under the dualization from gauge to height variables. With open boundary conditions, each configuration of gauge fluxes is equivalent to two height variable configurations.}
    \label{fig:hexafoillattice_dualization}
\end{figure}

\paragraph*{Model} The pure gauge {\uone} theory on the hexafoil lattice is represented in~\Cref{fig:hexafoillattice_dualization}a) with the connectivity graph denoted by $\mathcal{G}_{\hexafoil} = (\mathcal{V}_{\hexafoil}, \mathcal{E}_{\hexafoil})$. Note that if the petal links are merged, then the lattice geometry becomes triangular (the $U(1)$ quantum link model on the triangular lattice has been studied in~\citep{Banerjee2022NematicConfinedPhasesa}). We work in the simplest QLM setting, with spin-$1/2$ variables carrying electric flux $E_l \in \{ \pm 1/2 \}$ assigned to each link $l \in \mathcal{E}_{\hexafoil}$ (blue links in the figure). Lattice links carry a standard orientation shown by the arrows in the figure. The links on the hexafoil lattice form elementary triangle plaquettes (the three-link loops in blue) and petal plaquettes (the two-link loops in blue). We choose the states $\ket{0}$ and $\ket{1}$ of the gauge links to represent fluxes circulating respectively in the same or opposite direction compared to the standard orientation, such that the electric field always satisfies $E_l = Z_l/2$, with raising operators $U_l = (X_l + i Y_l)/2$. If $E_l<0$ it is useful to imagine the corresponding arrow to be flipped compared to the standard orientation. We will restrict ourselves to the pure gauge sector of the theory with only static charges at the vertices $v \in \mathcal{V}_{\hexafoil}$ (blue intersections in the figure). 

For the spin-$1/2$ $\mathrm{U}(1)$ quantum link model, $E_l^2=1/4$ is constant, and can therefore be omitted from the Hamiltonian. On the hexafoil lattice, the set of triangle plaquettes ($T$) and petal plaquettes ($P$) can be assigned distinct coupling constants; normalizing the Hamiltonian such that the triangle coupling is $1$ gives the general form
\begin{equation}
\begin{aligned}\label{eq:hamiltonian_hexafoil}
    H &= - \sum_{t\in T} (U_{t} + U_{t}^\dag) - K_P \sum_{p \in P} (U_{p} + U_{p}^\dag) \ , \\
\end{aligned}
\end{equation}
where $U_{t} = \prod_{l\in \partial t} U_{l}$ and $U_{p} = \prod_{l\in \partial p} U_{l}$ are the clockwise products of electric field raising operators $U_l$ along the links at the boundary of each triangle and petal plaquette.
As shown in~\Cref{fig:hexafoillattice_dualization}b), the two terms of the Hamiltonian act on triangle and petal plaquettes, respectively, by annihilating the configurations without circulating flux or flipping the orientation of the circulating plaquettes. Plaquettes with a clockwise or anti-clockwise circulation of electric flux are termed ``flippable''.
In previous works on the {\uone} quantum link model~\citep{Banerjee201321dU1,Banerjee2022NematicConfinedPhasesa}, a Rokhsar-Kivelson term~\citep{Rokhsar1988SuperconductivityQuantumHardCorea} was added to the Hamiltonian. In this case, we only discuss this possibility briefly in~\Cref{sec:rokhsar}, but we note that it does not lead to qualitatively different physics and is therefore not discussed in the main text. The Hamiltonian commutes with the local \uone~gauge symmetries generated by
\begin{equation}
    G_v = \sum_{l \in \mathrm{out}(v)} E_{l} - \sum_{l \in \mathrm{in}(v)} E_{l}\, , \quad\forall v\in \mathcal{V}_{\hexafoil}\, ,
\end{equation}
where $\mathrm{out}(v)$ and $\mathrm{in}(v)$ respectively indicate the links for which
their standard orientation is outgoing or ingoing from site $v$. These symmetries represent the conservation of flux at each node of the physical hexafoil lattice, with integer-valued eigenvalues $g_v$ that can be interpreted as external static charges.

\paragraph*{Dual representation} 
In the absence of dynamical matter, the model admits a dual description in terms of spin-$1/2$ \emph{height variables} placed on a dual graph $\mathcal{G}_{\heavyhex} = (\mathcal{V}_{\heavyhex}, \mathcal{E}_{\heavyhex})$ with vertices at the center of the plaquettes of the hexafoil lattice (see the black nodes and edges in~\Cref{fig:hexafoillattice_dualization}). In this way a local two-dimensional Hilbert space is associated with each dual vertex, corresponding to either a petal or triangle of the original graph. Importantly, the topology of this dual representation matches the ``heavy hex'' connectivity graph of the IBM quantum computers, which are the experimental quantum platform used in this work~\citep{Kim2023EvidenceUtilityQuantumb}. The relevant Pauli operators acting on the dual vertices associated with petals $p$ and triangles $t$ are written as $X_p$, $Y_p$, $Z_p$ and $X_t$, $Y_t$, $Z_t$. We make the conventional choice to define the height variables as $h_{p}= -Z_{p}/2 \in \{-\tfrac{1}{2}, \tfrac{1}{2}\}$  and $h_{t}= 1/2 - Z_{t}/2 \in \{0, 1\}$.
Each edge $l\in \mathcal{G}_{\hexafoil}$ of the original lattice is adjacent to one petal plaquette $p(l)$ and one triangle plaquette $t(l)$ of the dual lattice, and their associated height variables can be directly related to the original degrees of freedom via either of the equivalent relations
\begin{equation}\label{eq:dualization_relation}
\begin{cases}
    E_l = (h_{t(l)} - h_{p(l)})~\mathrm{mod}[2]~, & \forall l \in \mathcal{E}_{\hexafoil} \, ,\\
    Z_l = Z_{t(l)} Z_{p(l)}~, & \forall l \in \mathcal{E}_{\hexafoil} \, .
\end{cases}
\end{equation}
The first formulation, in particular, is useful to show that the height variable representation natively enforces the Gauss law modulo two in the bulk of the lattice, in the sense that for vertices $v\in \mathcal{V}_{\hexafoil}$
\begin{equation}\label{eq:gauss_law_native}
    G_v~\mathrm{mod}[2] =  \sum_{l \in \text{edges}(v)} (h_{t(l)} - h_{p(l)})~\mathrm{mod}[2] = 0\, .
\end{equation}
This partial enforcement of the gauge constraints reduces drift from the physical gauge sector under time evolution. However, the remaining Gauss law violations, specifically those involving even charges, must still be treated explicitly.
Moreover, due to~\Cref{eq:gauss_law_native}, only even static charges can be straightforwardly introduced in the height basis by choosing an appropriate initial state. Odd static charges (such as $q = \pm 1$) instead require a modification of the theory via the introduction of a \emph{Dirac string} connecting the two charges. Alternatively, with open boundary conditions, odd static charges may be placed on the boundary.

The second identity, in turn, can be used to derive the dual Hamiltonian of the $\mathrm{U}(1)$ quantum link model in \Cref{eq:hamiltonian_hexafoil}
\begin{equation}\label{eq:hamiltonian_heavyhex}
\begin{gathered}
    H_\mathrm{dual}(K_P) = H_T + K_P H_P, \\
    H_T = - \sum_{t\in T} \mathbb{P}_t X_{t}, \quad H_P = -\sum_{p \in P} \mathbb{P}_p X_{p}\, ,
\end{gathered}
\end{equation}
where we have introduced projection operators $\mathbb{P}_{t/p}$ onto flippable plaquettes and the sums run over the respective sublattices $T$ and $P$ corresponding to triangle and petal variables in the dual lattice.
As already mentioned, the triangle and petal Hamiltonians act by flipping the corresponding plaquettes and annihilating non-flippable plaquettes. This last constraint is implemented in the dual formulation by $\mathbb{P}_{t/p}$, which equal one if the corresponding plaquette is flippable and zero otherwise; therefore they can also be used to measure the flippability of given plaquettes.
In terms of the height variables, these operators $\mathbb{P}_{t/p}$ act on the nearest-neighbors of the site $t$ (resp.\ $p$), with respect to the heavy-hexagonal lattice, i.e.
\begin{equation}\label{eq:flippability operators}
\begin{gathered}
    \mathbb{P}_t = \ket{000}\bra{000} + \ket{111}\bra{111} \ ,\\
    \mathbb{P}_p = \ket{00}\bra{00} + \ket{11}\bra{11} \ ,
\end{gathered}
\end{equation}
where we note that triangles have three neighbours while petals have only two. In the dual height basis the flippability of a plaquette is therefore determined by its neighboring height variables in the other sublattice. The flippability operators~\Cref{eq:flippability operators} are the main observables we will measure. Non-zero values for the flippabilities of the triangle (resp.\ petal) correspond to the neighboring petal
(resp.\ triangle) height variables being ordered.
Similar operators have been used extensively in a single dimension or in the context of Rydberg atoms to define spin Hamiltonians with local symmetries~\citep{Surace2020LatticeGaugeTheoriesa,Michailidis2020SlowQuantumThermalizationb,Michailidis2020StabilizingTwodimensionalQuantumb,Yue2DPXP}. 

All correspondences between states and operators under the dualization are depicted in~\Cref{fig:hexafoillattice_dualization}b). Note that, even in the Gauss law sectors with even static charges, the mapping to height variables is not a one-to-one correspondence. In fact, each gauge link configuration corresponds to two equivalent height variable configurations related by a global shift leaving the electric field variables invariant. Specifically, the global shift
\begin{equation}\label{eq:global redundancy}
    h_x \to h_x +1 \quad \mathrm{mod}[2] \ ,
\end{equation}
flips all the height variables, but leaves the electric fields invariant, as is clear from~\Cref{eq:dualization_relation}.
Following the constraints of the physical hardware platform, we work with open boundary conditions (OBC) and fix this extra unphysical redundancy by a specific choice of boundary condition on the height variables. We note that periodic boundary conditions of the original lattice correspond to a sum over winding boundary conditions for the dual Hamiltonian~\labelcref{eq:hamiltonian_heavyhex}, but with open boundaries the correspondence between the original Hamiltonian~\labelcref{eq:hamiltonian_hexafoil} and its dual~\labelcref{eq:hamiltonian_heavyhex} is exact, and therefore results from the two formulations can be directly compared. Additionally, OBCs allow us to insert odd static charges at the boundary of the system without contradicting \Cref{eq:gauss_law_native}.

\paragraph*{Symmetries}
In the infinite-volume limit, the hexafoil lattice is symmetric under rotations and reflections that map the lattice onto itself. These form the dihedral group $D_6$ of order twelve. In finite volume, depending on the chosen arrangement, the symmetry group may be broken to a subgroup of $D_6$. Similarly, the Hamiltonian is invariant under lattice translations in infinite volume, although some or all of these may be broken in finite volume or by the open boundaries.

The Hamiltonian in~\Cref{eq:hamiltonian_hexafoil} is also invariant under a charge conjugation symmetry which acts as $C: U_l \to U_l^\dagger$ and $C: E_l \to -E_l$. In the dual theory, this is realized by either flipping all petals $C: h_p \to -h_p$ or all triangles $C: h_t \to 1-h_t$; both transformations are equivalent up to the global redundancy in~\Cref{eq:global redundancy}.

The model also admits the following $\mathbb{Z}_2$ unitary transformation: $\mathcal{S}_P = \prod_{p \in P} Z_p$, which flips the sign of the petal kinetic coupling:
\begin{equation} \label{eq:phaseFlip}
    \mathcal{S}_P~H_\mathrm{dual}(K_P)~\mathcal{S}_P = H_\mathrm{dual}(-K_P) \ ,
\end{equation}
Because of this relation, we can restrict ourselves to $K_P \geq 0$.

Moreover, this model also admits a chiral transformation given by the following operator in the flux basis: $\Gamma = \prod_{l \in C}(2 E_l)$, where $C \subset \mathcal{E}_{\hexafoil}$ is chosen such that every triangle and every petal contains exactly one link from $C$. Then, both plaquette terms change sign under the action of $\Gamma$, leading to
\begin{equation}
    \Gamma H (K_P) \Gamma = -H (K_P) \ .
\end{equation}
In particular this implies that the spectrum of the Hamiltonian is symmetric around zero.

Finally, we note that further conserved quantities exist related to winding symmetries. With periodic boundary conditions, the total electric flux
\begin{equation}
    E_{\mathrm{tot},\gamma} = \sum_{l \in \gamma} E_l
\end{equation}
along a non-contractible loops $\gamma$ in the dual lattice is conserved.
With open boundaries, most relevant for quantum simulation experiments, the same quantity is conserved for dual-lattice paths $\gamma$ that connect different points on the boundary.

\subsection{Phase diagram in the absence of static charges}
\label{sec:phase_diagram_qmc}
The phase diagram of the (2+1)D {\uone} quantum link model with $s=1/2$ has been studied on the square~\citep{Banerjee201321dU1} and triangular~\citep{Banerjee2022NematicConfinedPhasesa} lattices with a cluster quantum Monte Carlo (QMC) algorithm~\citep{Banerjee2021RecentProgressCluster, Banerjee2018SU2QuantumLink}. In the present work, a similar algorithm has been developed for the hexafoil lattice to study the phase diagram with this geometry; details can be found in~\Cref{sec:appendix_qmc-algorithm}. The algorithm works directly in the dual height formulation.

The phase diagram is characterized by order parameters such as the average magnetizations $M_T$ and $M_P$ of the triangle and petal height variables,  
\begin{equation}
    M_T = \frac{2}{V_T}\sum_{t \in T} \left( h_t-\tfrac12 \right) \ ,\qquad M_P = \frac{2}{V_P}\sum_{p \in P} h_p \ ,
\end{equation}
where $V_{T/P}$ are the number of triangles and petal height variables. These definitions ensure that $M_T, M_P \in [-1, 1]$. Note that because of the global ambiguity in the definition of the height variables, physically one should identify $(M_T, M_P) \sim (-M_T, -M_P)$. 
A non-zero expectation value of either observable indicates the ordering of the corresponding height variable and would thus signal the breaking of charge conjugation.

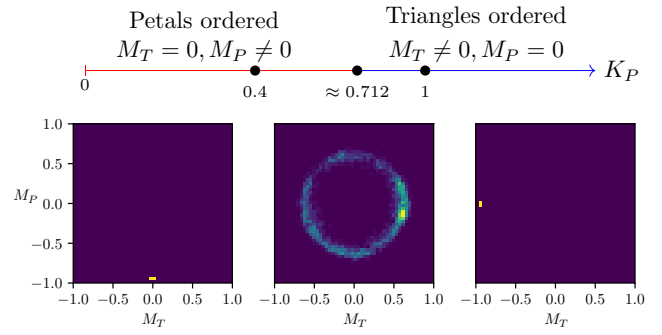
\begin{figure}[ht!]
    \centering
    \hfill
    \scalebox{0.9}{
    \begin{tikzpicture}

    \draw[|-,red] (0.5,0) -- (4.5,0) {};
    \draw[->,blue] (4.5,0) -- (8,0) node[right] {\color{black} $K_P$};

    \node[below] at (0.5,0) {\scriptsize  $0$};
    \draw (3,0) node[circle,fill,inner sep=1.5pt]{};
    \node[below] at (3,-0.1) {\scriptsize  $0.4$};
    
    \draw (4.5,0) node[circle,fill,inner sep=1.5pt]{};
    \node[below] at (4.5,-0.1) {\scriptsize $\approx 0.712$};
    \draw (5.5,0) node[circle,fill,inner sep=1.5pt]{};
    \node[below] at (5.5,-0.1) {\scriptsize $1$};

    \node[above] at (6.25,0.5) {Triangles ordered};
    \node[above] at (2.25,0.5) {Petals ordered};
    \node[above] at (6.25,0) {$M_T\neq 0, M_P=0$};
    \node[above] at (2.25,0) {$M_T= 0, M_P\neq 0$};
    \end{tikzpicture}
    }

    \includegraphics[width=\linewidth]{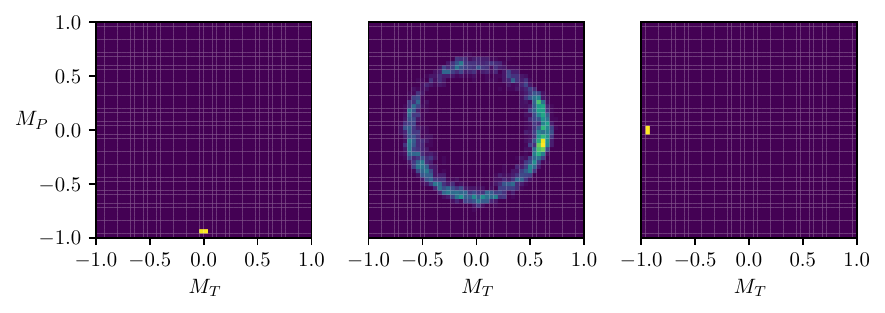}
    \caption{Histograms of the sampled Monte Carlo distributions of the $M_T, M_P$ observables at three different values of $K_P$, namely $K_P=0.4$ (phase where the petals are ordered), $K_P=0.712$ (the transition point) and $K_P=1.0$ (phase where the triangles order). The central picture shows the emergent {\uone} symmetry at the transition point.
    }
    \label{fig:MT_MP_history_combined}
\end{figure}

A sketch of the phase diagram, as $K_P$ is varied, is given in~\Cref{fig:MT_MP_history_combined}. This follows from the results of QMC simulations on a lattice with $48$ rows and columns of hexagons with periodic boundary conditions.
Histograms of $M_T$ and $M_P$ are shown for a representative set of $K_P$ values in~\Cref{fig:MT_MP_history_combined}. Away from the transition point, one of the two order parameters takes on a non-zero expectation value.
In particular, for small values of $K_P$, the petals are ordered while the triangles are disordered; for large $K_P$, the triangles are ordered, while the petals are disordered. In both phases, charge conjugation is broken, but no further symmetry distinguishes the two phases.
The lattice is large enough that tunneling between different vacua is strongly suppressed, so that only one of the two equivalent values of the non-zero order parameter is explored in either broken-symmetry phase. Although less sharp, these basic features of the phase diagram are already visible via exact diagonalization studies on a single hexagon, as discussed in~\Cref{sec:appendix_exact_diagonalization}. It should be noted that the QMC algorithm is inefficient for large values of $K_P$, but both the quantum simulation and exact diagonalization studies confirm the picture of the phase diagram given in~\Cref{fig:MT_MP_history_combined}.

The physics at the transition point between the two phases of the model is especially interesting. For a lattice with $48$ rows and columns of hexagons, the transition point is located at $K_P=0.712(1)$. Of course, especially for small lattices, the exact location of the transition is dependent on the geometry and boundary conditions. Therefore one should expect the transition to be located roughly, but not exactly, at this point on the smaller geometries used in the quantum simulation experiments. The two phases coexist at the transition point, and one observes the emergence of a {\uone} symmetry, as indicated from a plot of the Monte Carlo histograms of the observables in the $(M_T, M_P)$-plane shown in~\Cref{fig:MT_MP_history_combined}. As had been previously observed~\citep{Banerjee201321dU1, Zhao2019SymmetryenhancedDiscontinuousPhase}, these results are similar to the phenomenon known as deconfined quantum criticality~\citep{Senthil2004DeconfinedQuantumCritical,Senthil2004QuantumCriticalityLandauGinzburgWilson}.  
In this model, however, as we will see, the string tension becomes small near the transition point but does not vanish.
\section{Quantum quench across static phase transitions} \label{sec:quench}
The Hamiltonian $H_\mathrm{dual}(K_P)$ features string-like excitations connecting individual static charges that can be studied as dynamical objects. In this section, we run experiments on IBM's \texttt{ibm\_pittsburgh} quantum computer to probe the pre-thermal dynamics of such string states under global quantum quenches. Initial string states with fixed endpoints can be prepared as simple product states on quantum hardware, which correspond to eigenstates in the limit ${K_P=\infty}$. We subject these initial string configurations to global quenches via real-time evolution under the parametrized Hamiltonian $H_\mathrm{dual}(K_P)$ for several finite values of $K_P$. During this evolution, strings can oscillate due to the plaquette interaction, but cannot break as the Hamiltonian does not contain dynamical matter. 
\begin{figure*}[ht!]
    \centering
    \includegraphics[width=\linewidth]{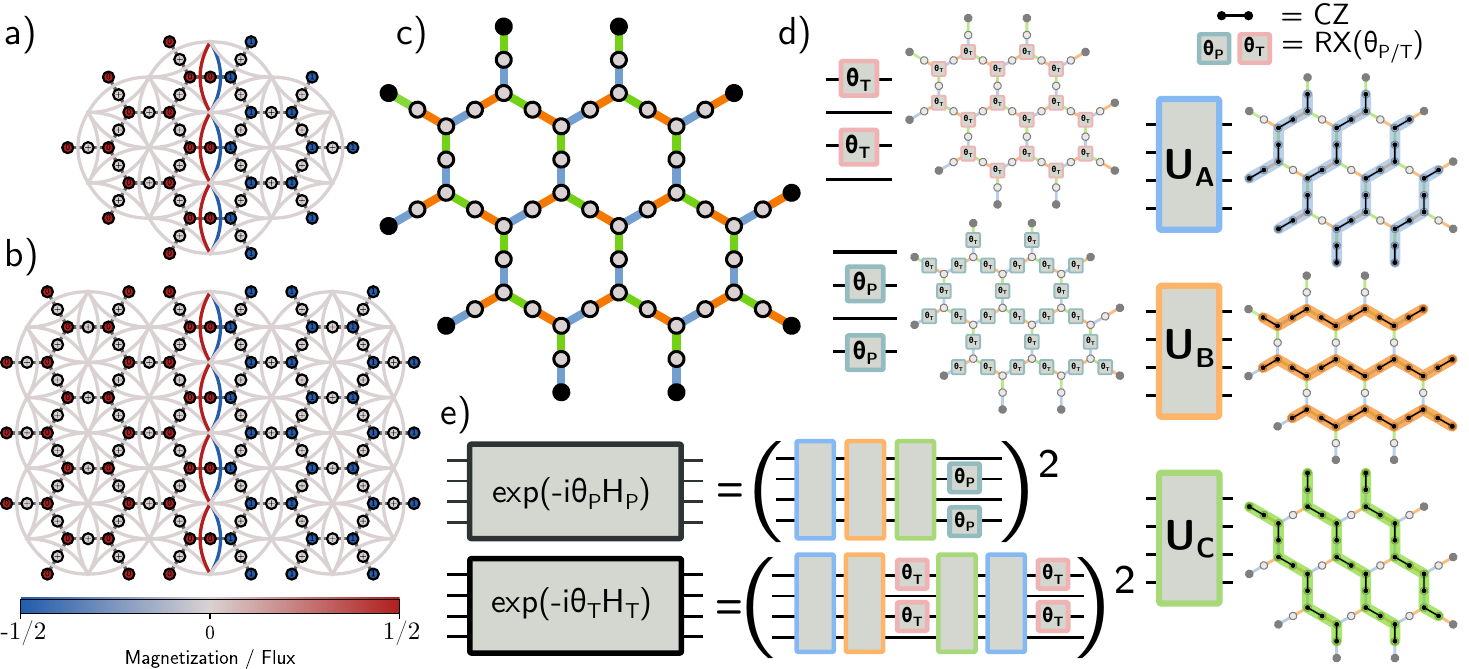}
    \caption{a)~b) Two sublattices of the heavy-hex hardware graph used in this work, with 4 and 13 heavy-hexagons respectively. The height variables are initially prepared in the lowest-energy eigenstate of $H_\mathrm{dual}(\infty)$ corresponding to a Gauss law sector with two static charges $q_v=\pm 1$ on the boundary. The initial qubit states are labeled as $\ket{0/1}$ and $\ket{\pm}$ with associated magnetization colored from blue to red. Trotter decomposition of the dual spin Hamiltonian $H_\mathrm{dual}$. c) Tricoloring of the edges $\mathcal{E}_{\heavyhex}$. d) Independent non-Clifford layers (pink and turquoise) and Clifford layers (blue, orange, green). e) This Trotter decomposition involves 6 entangling layers for the petal Hamiltonian $H_P$ and 8 entangling layers for the triangle Hamiltonian $H_T$.}
    \label{fig:trotterization}
\end{figure*}

\subsection{Quench dynamics}
We repeat the global quench experiments for two lattice geometries: one consisting of 4 full heavy-hexagons and the other of 13 full heavy-hexagons. For each geometry we perform experiments for coupling values $K_P \in [0.4, 0.7, 2.0]$. The first and third coupling values were chosen deep in the triangle-ordered and petal-ordered phases, respectively, while the second value is roughly tuned to the static phase transition point on large lattices.
The two lattice geometries used in this work are depicted in~\Cref{fig:trotterization}a)~b) and correspond to:
\begin{itemize}
    \item[a)] a smaller hexafoil lattice with 16 triangle plaquettes and 29 petal plaquettes. This is mapped to a system of 45 height variables on a heavy-hex graph,
    \item[b)] a larger hexafoil lattice with 42 triangle plaquettes and 72 petal plaquettes. This is mapped to a system of 114 height variables on a heavy-hex graph.
\end{itemize}

\paragraph*{Initial string-state configuration}
For the purpose of our experiment, we choose unit-flux strings connecting two static charges, $g_v=+1$ and $g_v=-1$, placed on the bottom and top boundary of the lattice, respectively. In~\Cref{fig:trotterization}a) b), these result from a specific choice of open boundary conditions on the triangle height variables. The initial states, eigenstates of $H_\mathrm{dual}(\infty)$, are the product states represented in~\Cref{fig:trotterization}a) b). 

\paragraph*{Circuit representation of the unitary dynamics}
\label{sec:appendix_trotterization}

In addition to the Hamiltonian $H_\mathrm{dual}$ being native on the heavy-hexagonal lattice implemented on IBM's devices, we find that the corresponding unitary dynamics admit a dense brick-wall circuit representation that is well-suited for currently available quantum hardware. We digitize the unitary time-evolution under the dual Hamiltonian $H_\mathrm{dual}$~\labelcref{eq:hamiltonian_heavyhex} using Suzuki-Trotter formulas. As the dual Hamiltonian~\labelcref{eq:hamiltonian_heavyhex} decomposes into two sums of commuting operators, $H_P$ and $H_T$, it follows that the implementation of second-order Trotter formulas only represents a constant overhead compared to the first-order formulas. We only consider second-order expansions of the following form
\begin{align}
    U_{ST2}&(k \times dt) = \exp(-i dt/2 H_T) \exp(-i K_P dt H_P) \notag \\
    &\cdot[\exp(-i dt H_T) \exp(-i K_P dt H_P)]^{k-1} \notag \\
    &\cdot\exp(-i dt/2 H_T) \, .
\end{align}
The circuit representation of each term in this expansion is shown in~\Cref{fig:trotterization}e).

\paragraph*{Error mitigation of hardware noise}
Our experiments are run on IBM's \texttt{ibm\_pittsburgh} quantum computer, with reported median $CZ$ errors of $1.6\times 10^{-3}$, single-qubit errors of $3.4\times 10^{-4}$ and reported readout assignment errors $1.3\times 10^{-2}$ at the time of the experiments. To mitigate the influence of hardware noise, we run extra calibration experiments to learn and invert a zeroth-order approximation of the noise channel in Clifford perturbation theory (CPT), as detailed in~\citep{Gonzales2025QuantumErrorCorrectionb,Schuhmacher2025ObservationHadronScatteringa} and summarized in~\Cref{sec:appendix_em_cpt}. Although the sampling cost of error mitigation is known to scale exponentially with the system size for recovering unbiased estimators of local observables~\citep{Takagi2022FundamentalLimitsQuantuma,Takagi2022FundamentalLimitsQuantuma}, we are only concerned here with constructing estimators with reduced bias and allow for an increase in the statistical variance for a fixed number of shots sampled from the quantum computer. 
In all of the following, we report the statistical errors for both the raw results of the quantum hardware---which only includes single-qubit twirling~\citep{Flammia2020EfficientEstimationPaulib,vandenBerg2023ProbabilisticErrorCancellation} of the entangling gates and of the readout---and the mitigated results after the inversion of the learned noise channel. We further discuss the systematic errors of both estimators via comparison with tensor network simulations of the Trotterized dynamics. For this purpose, we use a projected-entangled pair state (PEPS) approximation~\citep{Verstraete2004ValencebondStatesQuantum} of the quantum wavefunction with bond dimension $\chi=16$ that we assume to be converged unless stated otherwise. We separately study limits of the finite bond-dimension PEPS approximation in~\Cref{sec:appendix_peps_limits}.

\subsection{Results}
\begin{figure*}[t!]
    \centering
    \includegraphics[width=1\linewidth]{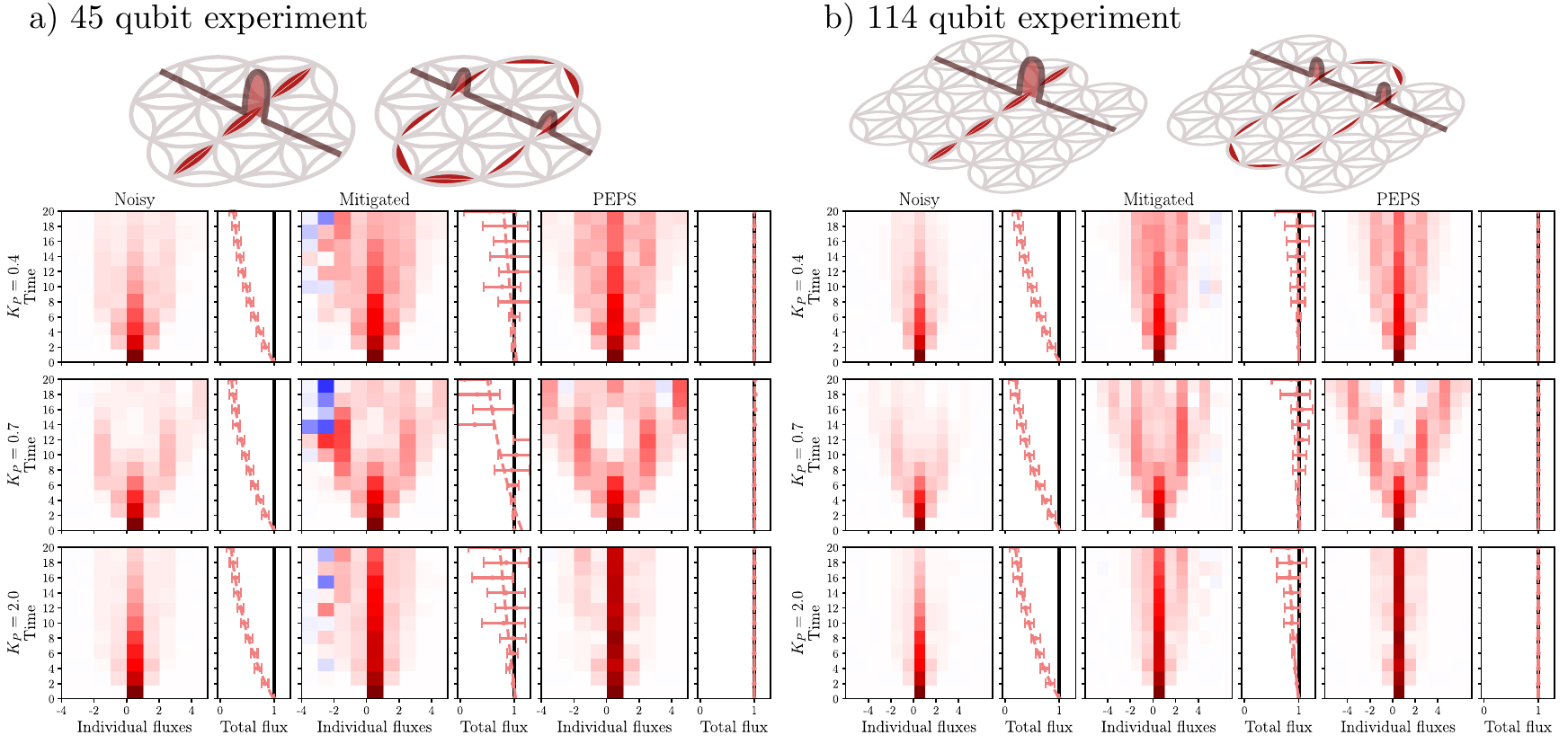} 
    \caption{Dynamics of the petal flux densities along a single cross-section for the lattices with respectively 4 and 13 full heavy-hexagons. The estimators from the quantum hardware, both raw and mitigated, are compared to reference PEPS results at multiple couplings $K_P\in\{0.4, 0.7, 2.0\}$. The single transversal slice, or cross-section, is represented at the top of the figure in two cartoon flux configurations that are explored throughout the dynamics. The first one corresponds to a narrow flux line carrying a full integer flux, while the second one corresponds to two well-resolved flux lines carrying half fluxes. The observables are measured for various circuit depths corresponding to total times $t =k\times dt$ with $k\in\{0,2,\dots,20\}$ and with a time step $dt=0.4$.}
    \label{fig:FIG04_flux_imshow}
\end{figure*}

\paragraph*{Transverse oscillations} By quenching an initial string state with the Hamiltonian $H(K_P)$, we probe the pre-thermal out-of-equilibrium dynamics. In the computational basis, the time-evolved state can be understood as a superposition of string states connecting the static charges $q=\pm 1$. The transverse oscillations of the initial flux string are captured by measuring the petal flux densities $\langle E_{l}-E_{l'} \rangle$, with $l$ and $l'$ the links composing a single petal, for all petals intersecting the cross-section
\begin{equation}
    \{-1\leq E_l-E_{l'} \leq 1,~~\forall p=(l,l')\in \text{cross-section}\}\, .
\end{equation}
In~\Cref{fig:FIG04_flux_imshow}, we show the petal flux expectation values extracted from the quantum hardware for time-evolution circuits up to total time $k\times dt=20\times 0.4 = 8$ (in units where the triangle coupling is $1$). We find good qualitative agreement between the raw hardware expectation values and the tensor network simulations of the same Trotter circuits up to around $15$  Trotter layers. At this point, the noise accumulated by the successive layers of entangling gates becomes too large and local expectation values approach the limits of a fully mixed state. The mitigated estimators, after inversion of an approximate noise channel, quantitatively agree with the reference results in all three regimes of interest.

At $K_P=2.0$, the initial state has a large overlap with the ground state of the quenched Hamiltonian. As a result, the time-evolved state quickly equilibrates to a steady state of the dynamics with a narrow flux string.
At $K_P=0.4$, we find a steady state of the dynamics where the flux connecting the charges spreads along a wider flux tube. At equilibrium, this corresponds to a reduced string tension which we further discuss in~\Cref{sec:static_phases}.
At $K_P=0.7$, the QMC results in equilibrium indicate an almost vanishing (but non-zero) string tension for the ground state. In our quench experiments, this translates into large transverse fluctuations of the initial string with linearly growing amplitude up to the final times accessible experimentally. In addition, the real-time dynamics show an oscillatory pattern between the first and second cartoon state depicted on top of~\Cref{fig:FIG04_flux_imshow}. At times $t\in \{10,12,14,16\}$, the time-evolved state has effectively two half-flux lines connecting the two charges and a net zero flux along the shortest-path string. This is reminiscent of the flux-fractionalization~\citep{Banerjee201321dU1,Banerjee2022NematicConfinedPhasesa,Banerjee2018SU2QuantumLink,Banerjee2024BrokenSymmetryFractionalized} observed in various QLMs in equilibrium, which we further present in~\Cref{sec:static_phases}.

\paragraph*{Flux conservation} As discussed in~\Cref{sec:model}, the gauge symmetries $G_v$ as well as the total fluxes $E_{\mathrm{tot},\gamma}$ across slices $\gamma$ orthogonal to the initial are invariant under the trotterized time-evolution.
For our choice of cross-section $\gamma$, represented schematically in~\Cref{fig:FIG04_flux_imshow} on both lattice geometries, we also represent the expectation values of the total flux $E_{\mathrm{tot},\gamma}$ in the side plots of~\Cref{fig:FIG04_flux_imshow}. These serve as additional probes of the quality of our experimental data. Across all regimes of our experiment, the mitigated estimators, although agnostic of the system's symmetries, are compatible with the expected conservation laws up to statistical uncertainties. This is a further indication of the applicability of CPT in this context. 

\begin{figure}[ht]
    \centering
    \includegraphics[width=1\linewidth]{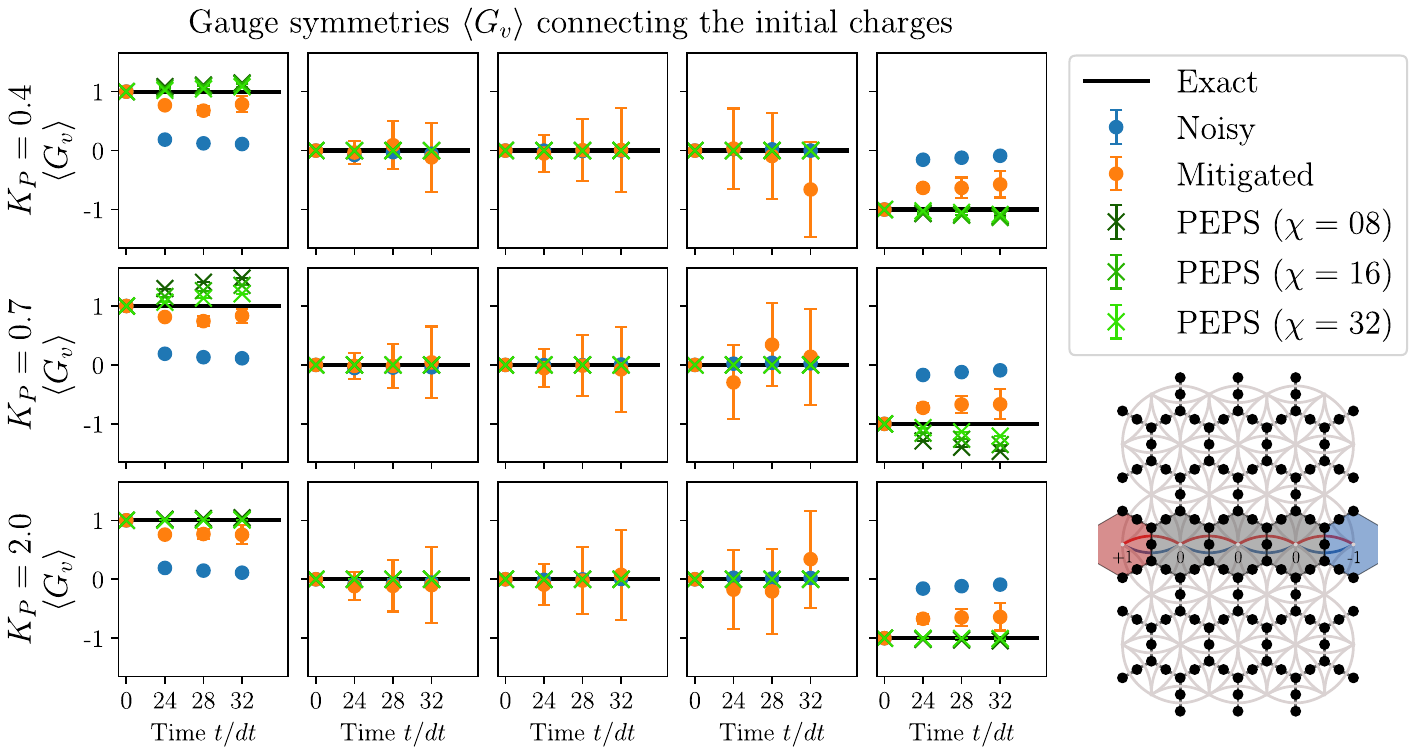}
    \caption{Expectation values of the gauge symmetry operators $G_v$ for the subset of sites $v\in \mathcal{V}_{\hexafoil}$ connecting the two static charges in the larger lattice with 13 heavy-hexagons. The drawing highlights the five gauge symmetries as well as their expected eigenvalues $g_v$. The experiment was repeated both on hardware (blue and orange points) and simulated with PEPS at finite bond-dimensions $\chi\in\{8,16,32\}$ (green points), at long times $t=k\times dt$ with $k\in\{24,28,32\}$ and for the couplings $K_P\in \{0.4, 0.7, 2.0\}$.}
    \label{fig:FIG10_symmetry_violations_v2}
\end{figure}

At longer times and with our implementation, the accumulation of hardware noise as well as the truncation to low bond-dimension PEPS both introduce violations of the gauge symmetries. The first is expected even for the mitigated signal due to the overhead associated with learning and inverting the noise channel, while the second is a result from fast-growing entanglement entropy in out-of-equilibrium dynamics. We present in~\Cref{fig:FIG10_symmetry_violations_v2} a comparison of the amplitudes of gauge violations for the five local symmetries $G_v$ connecting the initial static charges and at long time $k\times dt$ with $k\in\{24,28,32\}$. The complete data for all local symmetries is detailed in~\Cref{sec:appendix_peps_limits}. Noticeably, we find that, in the regime $K_P=0.7$, the deviations due to finite bond-dimension PEPS reach the same order-of-magnitude as those computed from the mitigation of our quantum hardware data. We hypothesize that this is related to a remarkable entanglement growth of the time-evolved state and indicates a regime of particular interest for the quantum simulation of our model.

\begin{figure}[ht!]
    \centering
    \includegraphics[width=1\linewidth]{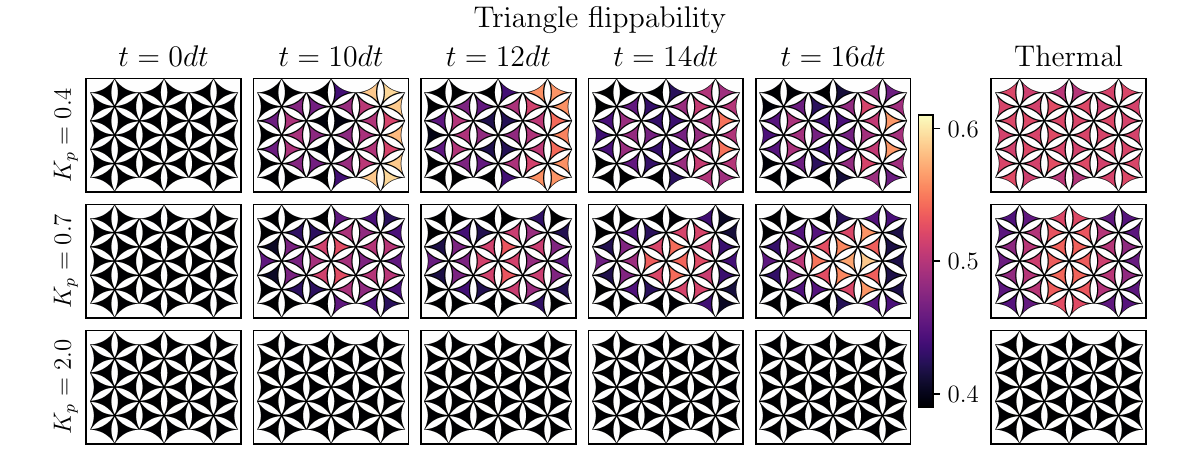}
    
    \vspace{0.3cm}
    \includegraphics[width=1\linewidth]{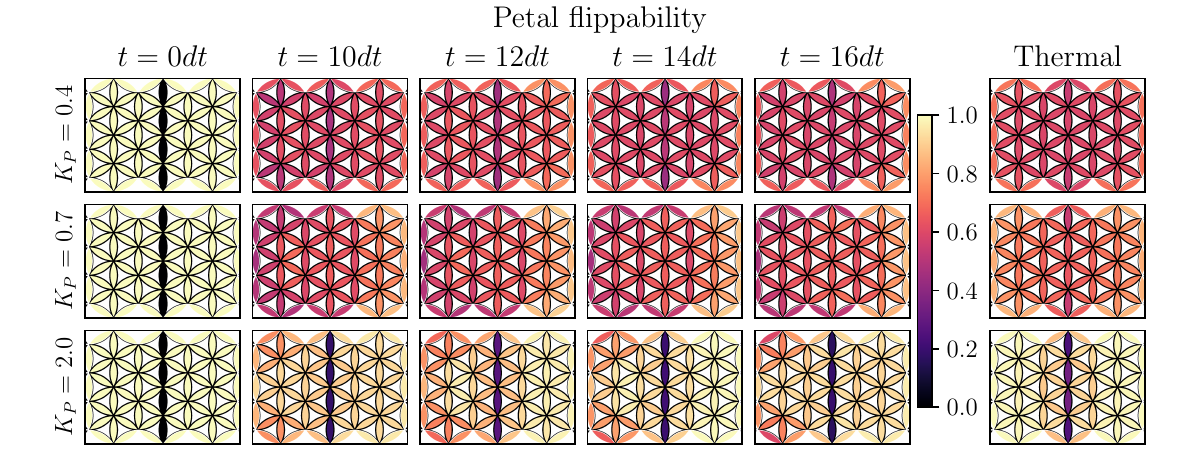}
    \caption{Local flippabilities for the three quenching regimes $K_P\in\{0.4, 0.7, 2.0\}$, and for a subset of measurement times $t = k\times dt$ with $k\in\{0,10,12,14,16\}$. At these times, the flux on the minimum-length line connecting the charges vanishes at the transition point $K_P=0.7$. Top) Flippabilities of triangle plaquettes $\mathbb{P}_t$. Bottom) Flippabilities of petal plaquettes $\mathbb{P}_p$. Values at thermal equilibrium (obtained by QMC) are also included. These correspond to the infinite time limit, assuming thermalization at the quench temperature given by the initial state configuration. Note that we rescaled the colors for the triangle flippabilities, although $\langle \mathbb{P}_{t} \rangle = 1/4$ in the initial state.}
    \label{fig:FIG05_flippabilities}
\end{figure}

\paragraph*{Flippabilities} During the quench dynamics at $K_P=0.7$, we find that the time-evolved state explores configurations schematically represented in~\Cref{fig:FIG04_flux_imshow} with two half-flux lines connecting the two integer charges. According to our understanding of flux-fractionalization in equilibrium, the flux lines connecting the static charges act as boundaries between the vacua of different phases. To probe the nature of the emergent phase in between the two half-flux lines, we measure local order parameter densities over time for the triangle and petal sublattices. In practice, both local height variables $h_p$, $h_t$ and local flippabilities $\mathbb{P}_{t/p}$ introduced in~\Cref{eq:flippability operators} can be used. Note that with our definition of the flippabilities, the non-zero values for the flippabilities of the triangle (resp.\ petal) correspond to the ordering of the neighboring petal (resp.\ triangle) height variables. Here we present our results for the flippabilities, which are insensitive to the redundancy of the height variable representations.
Direct measurements of the average height variables is also presented in \Cref{sec:appendix_detailed_hw}.
Our observations are depicted for the larger lattice of 13 heavy-hexagons and for various time slices in~\Cref{fig:FIG05_flippabilities}. One main feature of the density plots at $K_P=0.7$ is the large increase over time of the triangle flippability, corresponding to the petals ordering in the narrow band connecting the two static charges. The flux connecting the two charges in the time-evolved states is then channeled at the two boundaries of this region as seen in~\Cref{fig:FIG04_flux_imshow}.

\section{Confining properties in equilibrium} \label{sec:static_phases}

In addition to the phase diagram presented in~\Cref{sec:phase_diagram_qmc}, we have also performed numerical QMC simulations in equilibrium to understand the relation between the dynamics of the string width, monitored during the quench experiment, and the confining properties of the theory in equilibrium. From previous works with the \uone~quantum link model on different (2+1)D lattices~\citep{Banerjee201321dU1,Banerjee2022NematicConfinedPhasesa}, one expects that the theory is confining. In the presence of static charges, one can use QMC to measure various properties of the confining string and relate them to the dynamical properties observed in~\Cref{sec:quench}. 

\subsection{String tension in the hexafoil model}
Via QMC simulations in large volumes, we have computed the string tension for configurations with two $q=\pm 1$ static charges. The string tension $\sigma$ is extracted from the total energy $E(R)=\langle H \rangle$, in the presence of the two static charges separated by a distance $R$, by a fit to the phenomenological form
\begin{equation}
    \label{eq:phenomenological string energy}
    E(R) = A +\sigma R -\frac{C}{R} \ .
\end{equation}

\begin{figure}[ht!]
    \centering
    \includegraphics[width=\linewidth]{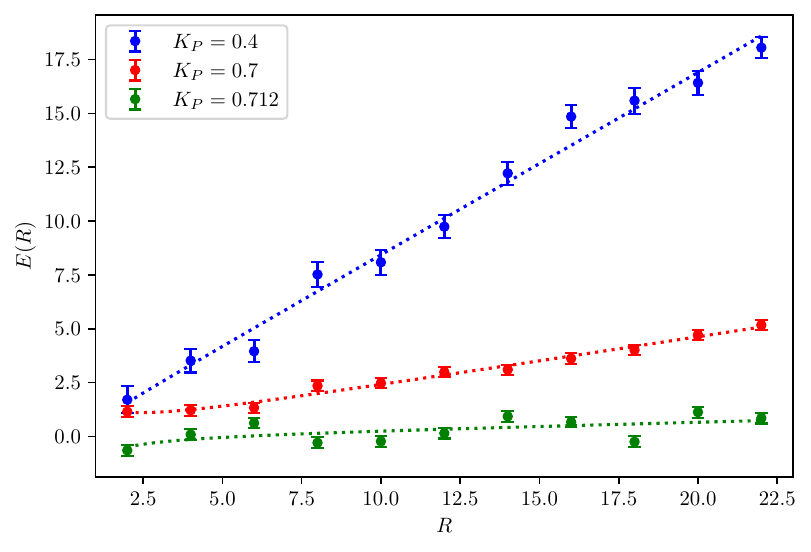}
    \caption{Energy of a static charge configuration ($q=\pm 1$) versus separation $R$ for various values of $K_P$. The data is well-described by a linearly rising potential, indicating confinement. Moreover, the string tension decreases as one approaches the phase transition.}
    \label{fig:E(R) vs R}
\end{figure}

We find that the string tension between $q=\pm 1$ charges is always non-zero. 
It is well-known that confining strings in gauge theories are usually well-described by effective string theory, which would predict a more stringent form for $E(R)$~\citep{Luscher1980AnomaliesFreeLoop,Luscher1981SymmetrybreakingAspectsRougheningb,Polchinski1991EffectiveStringTheory,Aharony2013EffectiveTheoryLong, Brandt2016EffectiveStringDescription, Caselle2021EffectiveStringDescription}. However, the present model is closely related to (2+1)D $\mathrm{U}(1)$ gauge theory, whose string is known to exhibit exceptional properties~\citep{Caselle2015DifferentKindString, Aharony2025EffectiveStringsQED3, Caselle2025FiniteTemperatureGround}; moreover the strings in the quantum link model are known to exhibit exotic behavior such as fractionalization. For these reasons, it is not clear whether effective string theory predictions are applicable. In any case, a precise study of the confining string is not the focus of the present work and we were able to obtain good fits with the phenomenological form~\Cref{eq:phenomenological string energy}. 

\begin{table}[]
    \centering
    \begin{tabular}{c ccc}
        \hline\hline \\[-1ex]
        $K_P$ & $0.4$ & $0.7$ & $0.712$ \\[1ex]
        {\quad $\sigma$ \quad} & $0.85(4)$ & $0.23(2)$ & $0.04(2)$ \\[1ex]
         \hline\hline
    \end{tabular}
    \caption{Measured values of the string tension.}
    \label{tab:string tension values}
\end{table}

The measured values of the string tension are given in~\Cref{tab:string tension values}, in units where the length of a petal in the hexafoil lattice (from tip to tip, see~\Cref{fig:trotterization}) is set to 1. Because the QMC algorithm becomes inefficient for large values of $K_P$, we were unable to extract a value for the string tension at $K_P=2.0$. At $K_P=0.712$, the string tension is very small but non-zero, consistent with results on the square and triangular lattices.

\begin{figure}[ht!]
    \centering
    \includegraphics[width=\linewidth]{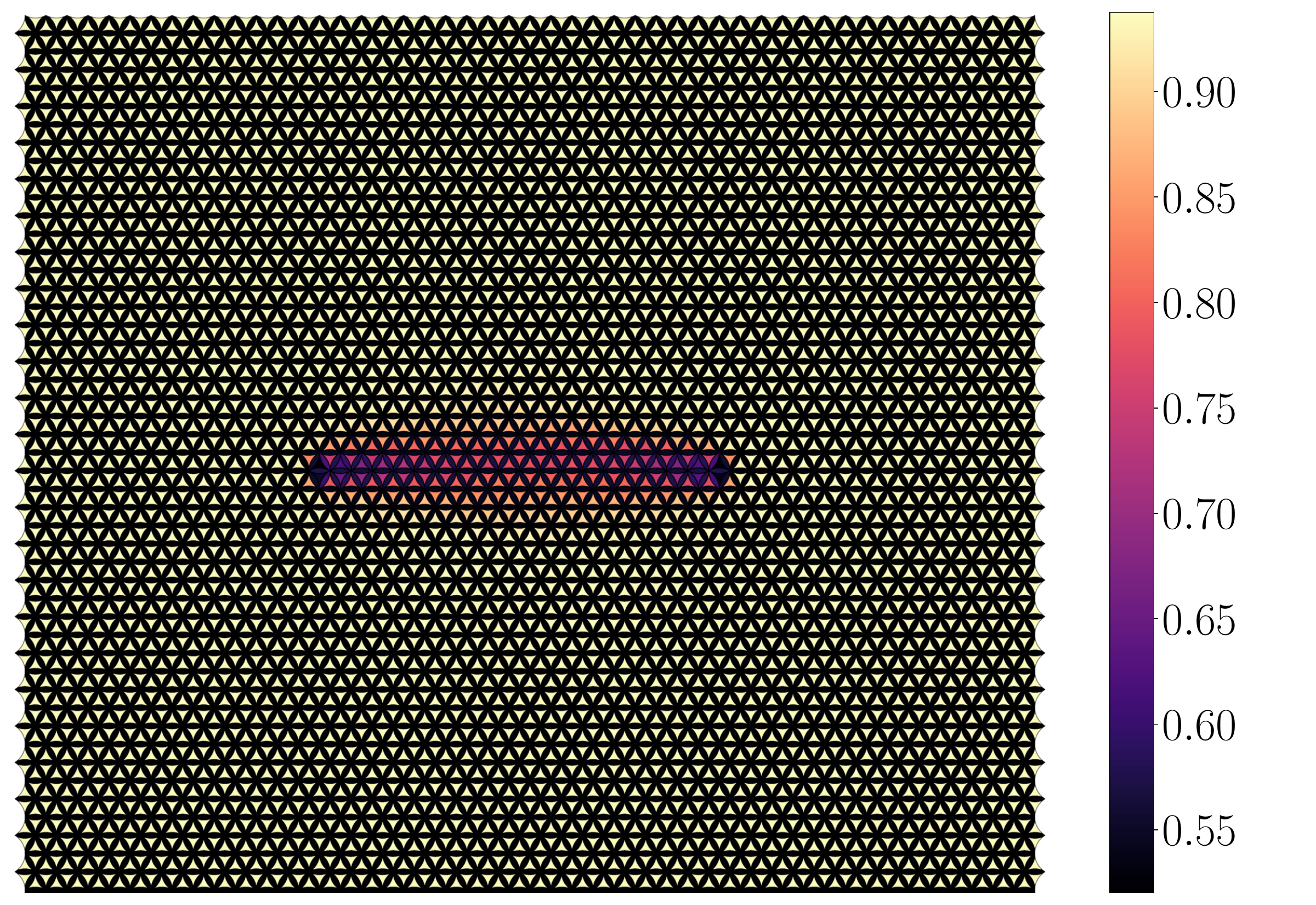}
    \caption{Local flippabilities of petals and triangles for a system of static charges ($q=\pm 1$) at $K_P=0.4$ and separation $R=20$.}
    \label{fig:string picture 0.4 flippability}
\end{figure}

\begin{figure}[ht!]
    \centering
    \includegraphics[width=\linewidth]{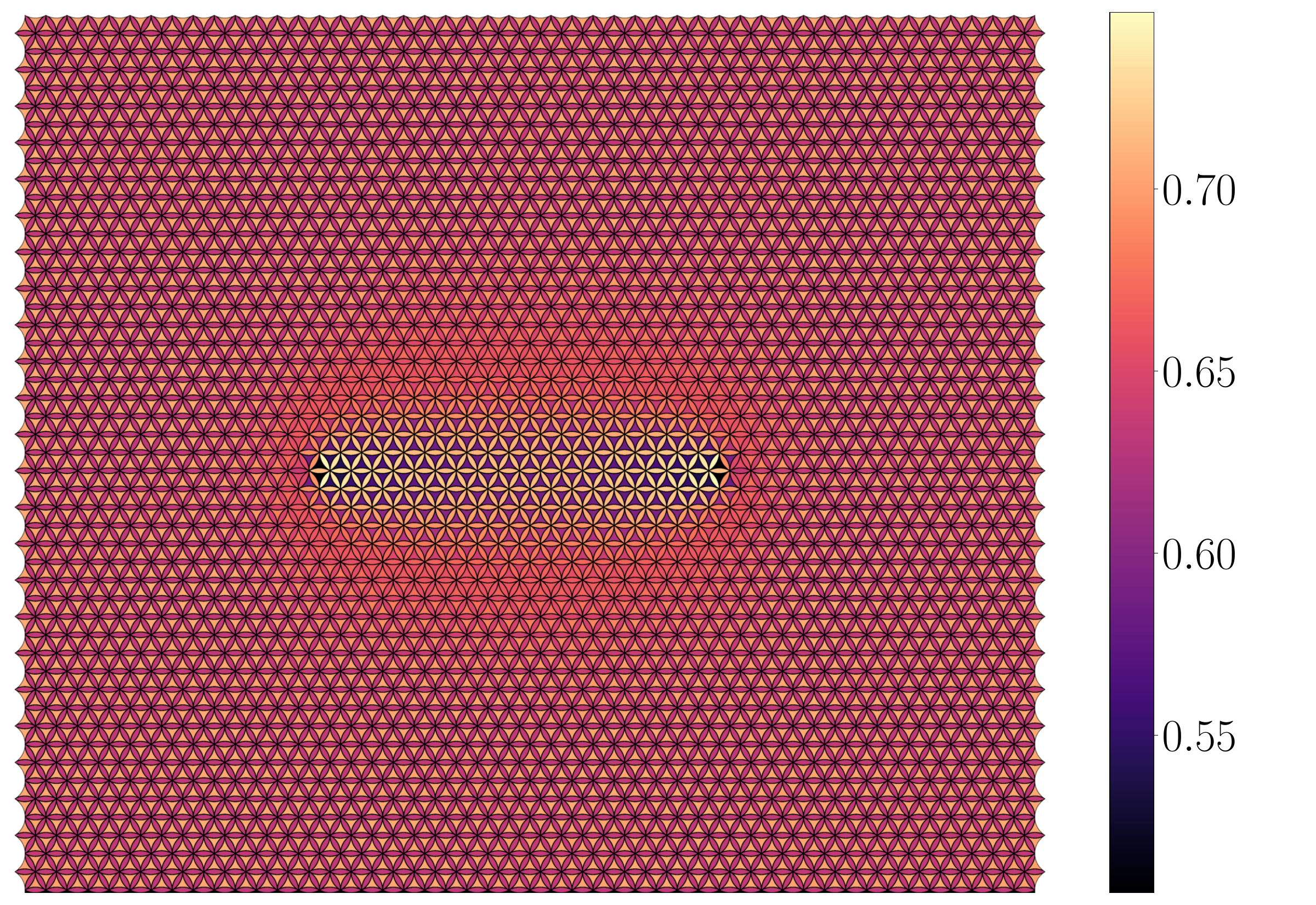}
    \caption{Local flippabilities of petals and triangles for a system of static charges ($q=\pm 1$) at $K_P=0.7$ and separation $R=20$. }
    \label{fig:string picture 0.7 flippability}
\end{figure}

Keeping in mind the caveats on the peculiarities of the string in this theory, from effective string theory one expects the mid-point width to scale as 
\begin{equation} \label{eq:string-width}
    w^2 \sim \frac{1}{\sigma} \log R \ .
\end{equation}
Therefore, as the string tension decreases, one expects the string width to increase, consistently with the values reported in~\cref{tab:string tension values} and the results of the quantum simulation experiments in~\Cref{fig:FIG04_flux_imshow}. For a qualitative observation of this phenomenon, \Cref{fig:string picture 0.4 flippability,fig:string picture 0.7 flippability} plot the local flippabilities as defined in~\Cref{eq:flippability operators} for equilibrium results at two choices of the coupling.
It is clear from the figures that the string is significantly wider at $K_P=0.7$ compared to $0.4$, with a ratio of approximately $3.5$ between the respective string widths, which is consistent with the previous discussion and expected scaling provided by~\Cref{eq:string-width} and~\Cref{tab:string tension values}.

It should be noted that from the figures, especially at $K_P=0.7$, it is not clear whether the string is fractionalized into multiple strands or is merely single-stranded and broadened. In contrast, our QMC studies performed at the $K_P = 0.712$ transition point demonstrate stronger signals of fractionalization, suggesting that the phenomenon may be narrowly localized near the phase transition. This situation is distinct from the square and triangular lattices, where fractionalization is observed in an entire region of the phase diagram~\citep{Banerjee201321dU1, Banerjee2022NematicConfinedPhasesa}. The fact that the real-time quantum simulation depicts signs of independent half-flux strands without fine-tuning to criticality is suggestive that the string fractionalization at criticality in the static theory may have a more significant presence in real-time evolution. The precise connection between these effects remains an interesting open question.

\section{Conclusions}

In this work we have considered a (2+1)D {\uone} quantum link model on an unusual, specifically tailored, lattice geometry such that its dual height variable formulation matches exactly the heavy-hex arrangement of the IBM quantum hardware used here. We have explored the properties of the confining string in this model, most importantly in a dynamical setting using quantum computation, but also in equilibrium using both Monte Carlo sampling and exact diagonalization. This study represents an example of how complementary quantum and classical techniques can be used to characterize the physical properties of a system.

From the point of view of real-time dynamics, the dual formulation yields an efficient quantum circuit representation of the digitized unitary evolution under the magnetic plaquette term, which is key to accessing genuine (2+1)D string dynamics~\citep{Tian2025RolePlaquetteTerma}. Additionally, the dual formulation partially solves Gauss' law, and one must then only take care of preventing even-charge Gauss law violations in the dynamics. Extra gauge-symmetry preserving terms can be added to the Hamiltonian, such as the Rokshar-Kivelson term discussed in~\Cref{sec:rokhsar}, and we leave their dynamical implementations in the dual formulation to future work. We ran multiple quantum hardware experiments of quantum quenches in our model with IBM's \texttt{ibm\_pittsburgh} device. These experiments investigate a state with external $q = \pm 1$ charges connected by a confining string.
Initializing to an eigenstate of the Hamiltonian $H_\text{dual}(K_P=\infty)$, the state is evolved under a Hamiltonian $H_\text{dual}(K_P)$ with finite couplings $K_P \in \{0.4, 0.7, 2.0\}$ representative respectively of points in the phase diagram where the triangles are ordered, the phase transition occurs, or the petals are ordered.
Quenching to the triangle-ordered phase shows a widening of the initial flux line connecting the two initial static charges. Additionally, quenching the system to the approximate transition point shows remarkable dynamical features even at early times. The flux connecting the charges is periodically channeled through two half-strands. In the interior of these strands, the time-evolved state is characterized by an increase of the triangle flippability, corresponding to a region where the petals order similarly to the vacuum at $K_P=0$. 

The experimental implementation of unitary evolution on quantum hardware is affected by multiple sources of error. To address this, we exploit the residual local symmetries of the dual Hamiltonian to validate the error-mitigated estimators constructed from hardware data. For the results presented in this work, we post-process the data using an expansion of the noise channel within the framework of Clifford perturbation theory~\citep{Farrell2024ScalableCircuitsPreparinga,Gonzales2025QuantumErrorCorrectionb,Schuhmacher2025ObservationHadronScatteringa}. 
We find good qualitative agreement of the mitigated estimators with tensor network references at early times, when the entanglement in the system is still small. At later times, the projection to the low bond-dimension PEPS manifold, as well as the hardware errors in the quantum simulation, induce symmetry violations which become probes of the quality of an estimator. Notably, we identify a regime near the phase transition where the symmetry violations associated with both methods become comparable in magnitude. The development of improved estimators capable of maintaining accuracy on quantum hardware at such long evolution times remains an important direction for future work.

Finally, we highlight the deep connections with condensed matter physics. On the square lattice, the pure gauge spin-$1/2$ QLM is equivalent to the quantum spin-ice model on the checkerboard lattice~\citep{Shannon2004CyclicExchangeIsolated,Sikora2011ExtendedQuantumU1liquid}. Our lattice geometry is formulated instead of a hexagonal grid, but our study can directly inspire similar works on kagome, maple leaf, or other exotic lattices relevant for condensed matter physics.

\section{Acknowledgements}
A.G. would like to thank Julian Schuhmacher for joint work on CPT, as well as Thea Budde, Joao C. Pinto Barros and Marina K. Marinkovi\'c for many discussions related to this work. The authors acknowledge the use of IBM Quantum Credits for this work and are grateful for the support of Sieglinde Pfaendler from the Credits Program. I.T., F.T, and A.G. thank the Swiss National Center of Competence in Research (NCCR) SPIN (funded by the Swiss National Science Foundation (SNSF) under grant number 180604) as well as the SNSF grant RESQUE project (grant number 225229) for financial support. D.B. would like to thank STFC (UK) consolidated grant ST/X000583/1 and continued support from the Alexander von Humboldt Foundation (Germany) in the context of the research fellowship for experienced researchers. A.M. is supported by the Simons Foundation grant 994300 (Simons Collaboration on Confinement and QCD Strings) as well as by the SFT Scientific Initiative of the Italian Nuclear Physics Institute (INFN). Some of the classical numerical simulations were run on CINECA computers. U.-J.W., G.K. and A.M. acknowledge funding from the Schweizerischer Nationalfonds (grant agreement number 200020\_200424). U.-J.W. acknowledges support from the Alexander von Humboldt Foundation (Germany) and is thankful for the hospitality at Bonn University on occasion of different visits

\bibliography{biblio_paper}

\clearpage
\newpage

\onecolumngrid
\appendix

\section{Long-time dynamics and limits of PEPS approximations}
\label{sec:appendix_peps_limits}
In the main text we compare the results from the quantum hardware to a converged tensor network reference based on Projected Entangled Pair states (PEPS). The convergence of the PEPS approximation holds at early times because the initial states are non-entangled, although we generally expect fast-growing entanglement entropy in out-of-equilibrium dynamics. Because entanglement entropy measures are not easily accessible on the quantum hardware within our error-mitigation scheme, we focus on indirect probes to the limits of the PEPS representation of the time-evolved quantum state. As discussed in~\Cref{sec:model}, the formulation of the pure gauge Hamiltonian in~\Cref{eq:hamiltonian_hexafoil} in terms of height variables solves Gauss' laws modulo 2. The remaining Gauss law violations, i.e., those with even charges, are only enforced digitally in the sense that unitary time evolution commutes with the local symmetries ${[U_{ST2}(k \times dt), G_v]=0}$. Note that our implementation of the real-time evolution assumes a digitized Trotter evolution even for the tensor networks and that we don't discuss the error induced by the second-order Trotter approximation used in this work. While the exact time-evolved quantum state shares the same symmetries as the initial state, the projection to the low bond-dimension PEPS manifold is not symmetry preserving in our implementation. As a result, we can use Gauss law violations $\| G_v - g_v \|^2$, with $\{g_v,~v\in \mathcal{V}_{\hexafoil}\}$ the static charges in the initial state, as local probes of the error induced by the finite bond-dimension constraint. 

\begin{figure}[!ht]
     \centering
     \includegraphics[width=0.3\linewidth]{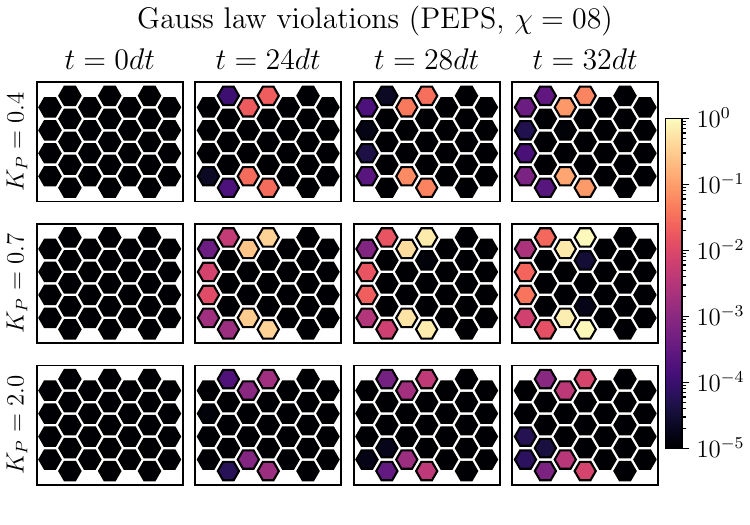}
     \includegraphics[width=0.3\linewidth]{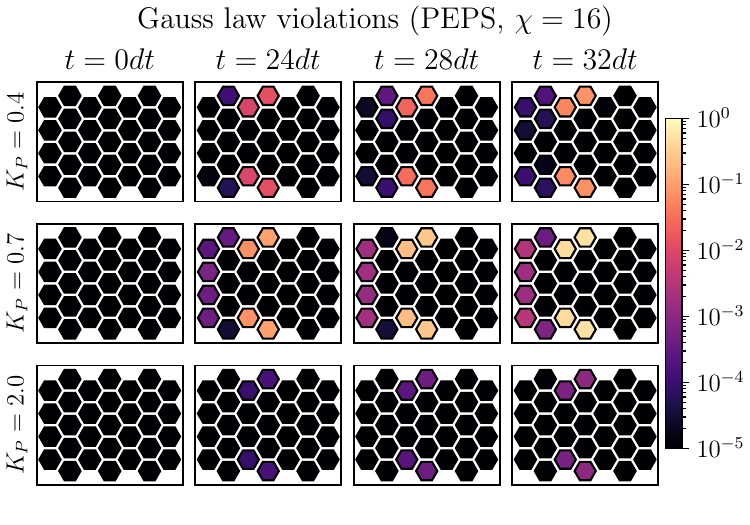}
     \includegraphics[width=0.3\linewidth]{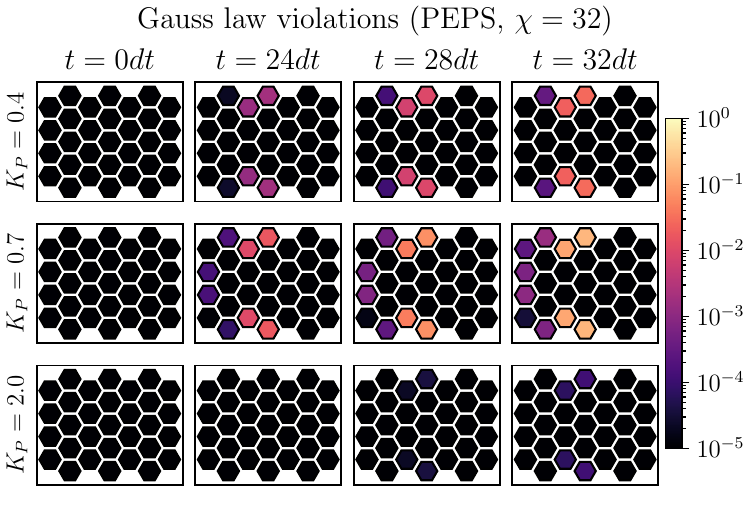}
     \caption{Local violations of Gauss' law for three PEPS approximations with bond-dimensions $\chi \in \{8, 16, 32\}$ and at long times $t=k\times dt$ with $k\in\{24, 28, 32\}$ and $dt=0.4$. There, the violations reach maximum values in the regime $K_P=0.7$ and in the vinicity of the initial static charges, with $\text{max}_{v}\| G_v - g_v \|^2 = \{0.22, 0.12, 0.04\}$ for respective bond dimensions $\chi \in \{8, 16, 32\}$.}
     \label{fig:limits_of_peps}
 \end{figure}

In this section, we present extended results for the gauge symmetry violations at long times $t=\{24, 28, 32\} \times dt$. We show in~\Cref{fig:limits_of_peps} the local symmetry violations for PEPS with bond-dimensions $\chi \in \{8, 16, 32\}$. The largest violations are generally found in the neighborhood of the initial static charges. For the smallest bond dimension $\chi=8$, these take a maximal value $\text{max}_{v}\| G_v - g_v \|^2 = 0.22$, which is of the same order-of-magnitude as the static charges themselves. These values are reduced to $\text{max}_{v}\| G_v - g_v \|^2 = 0.04$ for $\chi=32$, although with a large overhead in the classical compute time, as PEPS lacks efficient algorithms for contracting tensors~\citep{Lubasch2014UnifyingProjectedEntangled}.

\begin{figure}[!ht]
     \centering
     \includegraphics[width=0.3\linewidth]{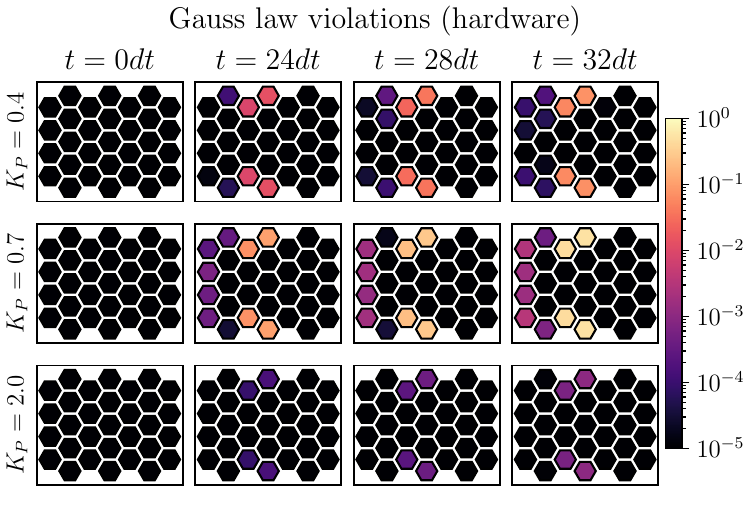}
     \caption{Local violations of Gauss law for the mitigated hardware data at long times $t=k\times dt$ with $k\in\{24, 28, 32\}$ and $dt=0.4$. The maximum values are observed at long times in the transition regime with $K_P=0.7$. There, the violations reach maximum values in the vinicity of the initial static charges, with $\text{max}_{v}\| G_v - g_v \|^2 = 0.12$ although with a large statistical variance.}
     \label{fig:limits_of_hardware}
 \end{figure}

On the other hand, the quantum hardware noise also induces a drift of the dynamics to a fully mixed state in the symmetry sector with no static charges. We ran additional experiments to compare the Gauss' law violations arising in the context of PEPS approximation and arising in our hardware experiments due to hardware noise. We report the expectation values $\langle G_v \rangle$ for the mitigated hardware results in ~\Cref{fig:limits_of_hardware}. These results are also compared with the tensor network approximations in~\Cref{fig:FIG10_symmetry_violations_v2} of the main text. Most notably for the present discussion, the Gauss' law violations are also maximal around the location of the initial static charges, and around the transition point $K_P=0.7$.

\section{Clifford perturbation theory expansion of the noise channel}
\label{sec:appendix_em_cpt}

When executed on physical quantum hardware, gates of the quantum circuit $U_{ST2}$ are executed with a finite fidelity that we model via a noise channel acting on top of the ideal unitary dynamics. The task of error-mitigation, \emph{i.e.} characterizing, learning, and inverting these noise channels is known to be associated with an exponential overhead in the sampling cost~\citep{Takagi2022FundamentalLimitsQuantuma,Takagi2023UniversalSamplingLowera}. In this work, we consider instead the approach of~\citep{Farrell2024ScalableCircuitsPreparinga,Gonzales2025QuantumErrorCorrectionb,Schuhmacher2025ObservationHadronScatteringa}, where the noise model of $U_{ST2}$ is approximated in Clifford perturbation theory (CPT) in the neighborhood of an observable $\hat{O}$ and inverted to construct a mitigated estimator $O^\mathrm{mitig}$. Notably, this estimator is \emph{not} unbiaised, and its sampling cost scales with the size of the neighborhood instead of the full system size. In the context of Trotter evolution with up to 30 Trotter steps, we report a good quantitative agreement of the mitigated estimators $O^\mathrm{mitig}$ of local symmetries with their exact values. Effectively, the CPT expansion captures relevant contributions to the noise in superconducting architectures, and appears to be a practical strategy for local observables, even far away from Clifford points ($\theta_{P/T}=0,\pi/2,...)$.
\\

\paragraph*{CPT expansion and noise learning}

The Trotter circuit $U_{ST2}$ is built from repeated applications of Clifford entangling layers and non-Clifford single-qubit rotations. As a direct consequence, setting the time step parameter $dt=0$ in these unitaries yields a Clifford unitary $U_{ST2}(k\times 0) = U_{T}(0)U_{P}(0) [U_{T}(0)U_{P}(0)U_{T}(0)]^{k-1} U_{P}(0)$ which acts as the identity on any initial state~$\rho_\mathrm{init}$. We assume, as is common for superconducting architectures, that the fidelity of two-qubit gates is one order of magnitude larger than that of single-qubit gates, such that the former can be safely neglected. This allows to use a local single-qubit twirling~\citep{vandenBerg2023ProbabilisticErrorCancellation} of the entangling gates and to obtain, for Clifford circuits, a global Pauli noise channel 
\begin{equation}\label{eq:pauli_global_noise_channel}
    \mathcal{E}_0(\rho_0) = \sum_{P_i\in \mathcal{P}^n} c_{P_i} P_i \rho_0 P_i,~~\text{with}~~ \rho_0 = U_{ST2}(0) \rho_\mathrm{init} U_{ST2}(0)^\dag = \rho_\mathrm{init}\, .
\end{equation}
If we are only interested in sampling the final quantum state in the computational basis, this general expression for a Pauli channel can be restricted to the subset of bitflip Pauli generators $P = \mathbf{X}^{b} = \bigotimes_i X_i^{b_i}$ that do affect the classical outcome of the measurement. The coefficients $c$ of a physical Pauli channel then represent a probability distribution $\mathbf{c} : \{0,1\}^n\mapsto [0,1]$. Now we can also express the global noise-channel $\mathcal{E}_{k\times dt}$ for the non-Clifford unitary $U_{ST2}(k\times dt)$, although it is in general non-Pauli. Clifford Perturbation Theory (CPT) suggests a practical expression of this channel as a Taylor expansion in the parameter $dt$
\begin{equation}
    \mathcal{E}_{k\times dt}(\cdot) = \mathcal{E}_{0}(\cdot) + dt (\partial_{dt} \mathcal{E}_{k\times dt})\big|_{dt=0}(\cdot) + \mathcal{O}(dt^2)\, .
\end{equation}

We now mitigate the zeroth-order contribution in the CPT expansion by applying the following steps:

\begin{itemize}
    \item \textbf{Raw estimator:} Collect the noisy samples from the quantum circuit implementing $\mathcal{E}_{k\times dt}(\rho_{k\times dt})$. The noisy samples are collected in a distribution $\tilde{\mathbf{s}}_{k\times dt}: \{0,1\}^n \mapsto [0,1]$. For a diagonal observable $O$ in the computational basis, we construct the raw estimator
    \begin{equation}
        \tilde{O}_\mathrm{raw} = \sum_{b_i\in \{0,1\}^n} [\tilde{\mathbf{s}}_{k\times dt}]_{b_i} \bra{b_i} O \ket{b_i}\, .
    \end{equation}


    \item \textbf{Noise characterization:} Learn the coefficients of the noise channel $\mathcal{E}_0$ from sampling the state $\mathcal{E}_0(\rho_0)$ on hardware. The noisy samples are collected in a distribution $\tilde{\mathbf{s}}_0: \{0,1\}^n \mapsto [0,1]$. The reference distribution $\mathbf{s}_0 = \mathbf{s}_\mathrm{init}$ is peaked and evaluates to $\delta(b_\mathrm{init})$, assuming $\rho_\mathrm{init}=\ket{b_\mathrm{init}}\bra{b_\mathrm{init}}$. Now both distributions are related by the Pauli noise channel in~\Cref{eq:pauli_global_noise_channel} via
    \begin{align}
        [\tilde{\mathbf{s}}_0]_{b_i} &= \mathrm{Tr}(\mathcal{E}_0(\rho_0) \ket{b_i}\bra{b_i}) \\
        &=\sum_{b_j} [\mathbf{c}]_{b_j}~\mathrm{Tr}(\mathbf{X}^{b_j} \rho_0 \mathbf{X}^{b_j} \ket{b_i}\bra{b_i}) \\
        &=\sum_{b_j} [\mathbf{c}]_{b_j}~[{\mathbf{s}}_0]_{b_i-b_j}\, ,
    \end{align}
    which is the standard convolution product of two distributions. One can solve this equality to estimate the unknown coefficients $[\tilde{\mathbf{c}}]$ by a fast Walsh-Hadamard transformation, under which the convolution product reduces to an element-wise product.
    \item \textbf{Zeroth-order CPT noise mitigated estimator:} The parameters of the noise channel can be used to construct an error-mitigated estimator $\tilde{O}_\mathrm{raw}$ for a diagonal observable $O$
    \begin{align}
        \tilde{O}^\mathrm{mitig} &= \mathrm{Tr}[\tilde{\mathcal{E}}_{0}^{-1}(\mathcal{E}_{k\times dt}(\rho_{k\times dt})) O] \\
        &= \sum_{b_i\in \{0,1\}^n} \left(\sum_{b_j\in \{0,1\}^n} [\tilde{\mathbf{c}}^{-1}]_{b_j}~[\tilde{\mathbf{s}}_{k\times dt}]_{b_i-b_j} \right) \bra{b_i} O \ket{b_i}\, ,
    \end{align}
    which is again a convolution product. The sampling cost of this mitigated estimator depends on the 1-norm of the inverse channel $\mathcal{E}_{0}^{-1}$~\citep{Bravyi2021MitigatingMeasurementErrorsa}, in this case the overhead factor is 
    \begin{equation}
        \gamma = \|[\tilde{\mathbf{c}}^{-1}]\|_1 = \|\mathrm{ifwht}(\mathrm{fwht}([\tilde{\mathbf{c}}])^{-1})\|_1\, .
    \end{equation}

\end{itemize}

\section{Detailed Hardware results}
\label{sec:appendix_detailed_hw}

\subsection{Dynamics of petal flux densities along a cross-section}
\begin{figure}[ht!]
    \centering
    \includegraphics[height=7cm]{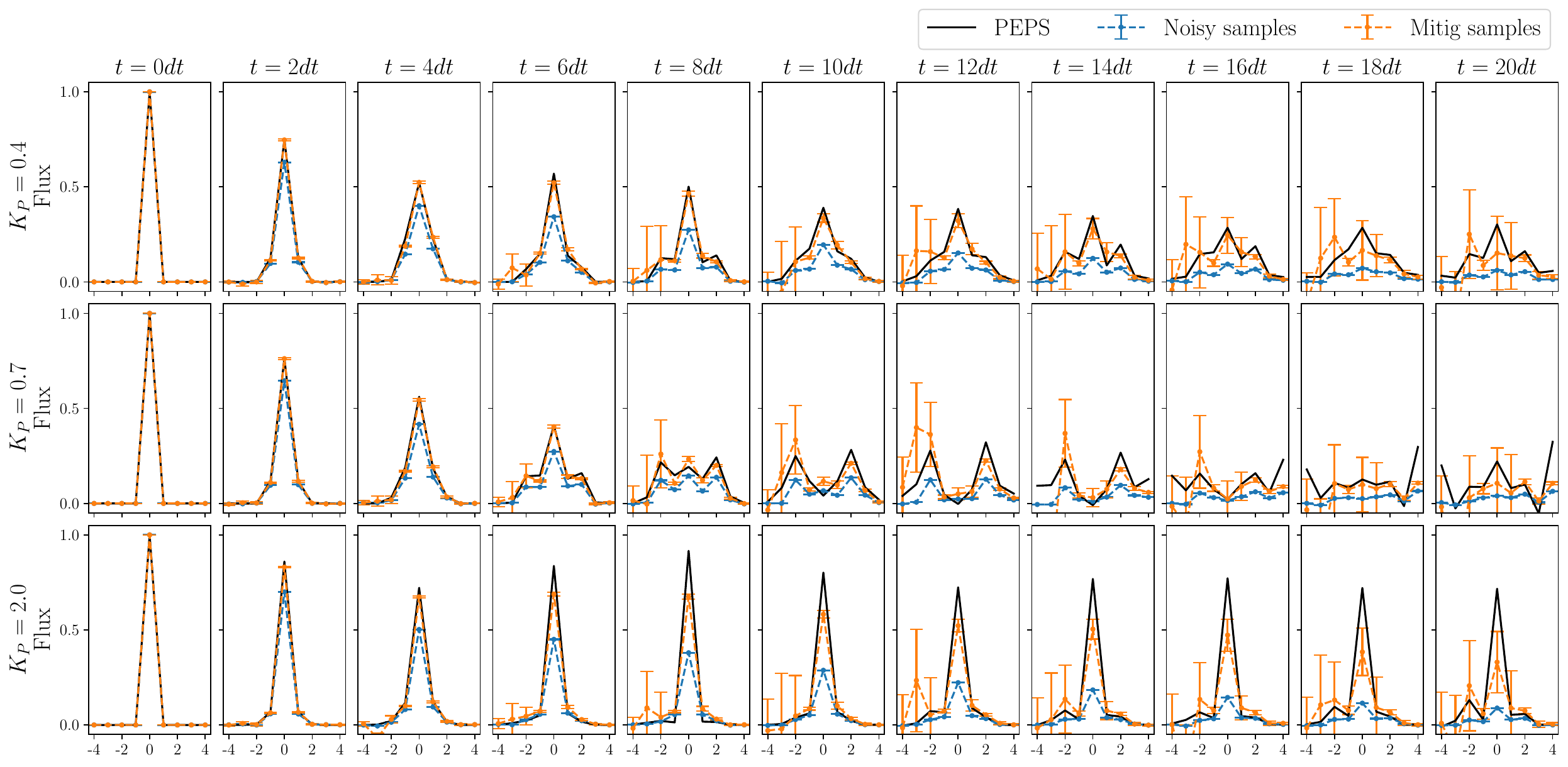}
    \caption{Dynamics of the petal flux density along the cross-section defined in~\Cref{fig:FIG04_flux_imshow} for the smaller lattice with 4 full heavy-hexagons, and for the three quantum quenches with couplings $K_P\in\{0.1, 0.7, 2.0\}$. The observables are measured for various circuit depths corresponding to total times $t=k\times dt$, $k\in\{0,2,\dots,20\}$, and to a time step $dt=0.4$ (in units where the triangle coupling is $1$).}
    \label{fig:FIG11_flux_imshow_bis_small}
\end{figure}

\begin{figure}[ht!]
    \includegraphics[height=7cm]{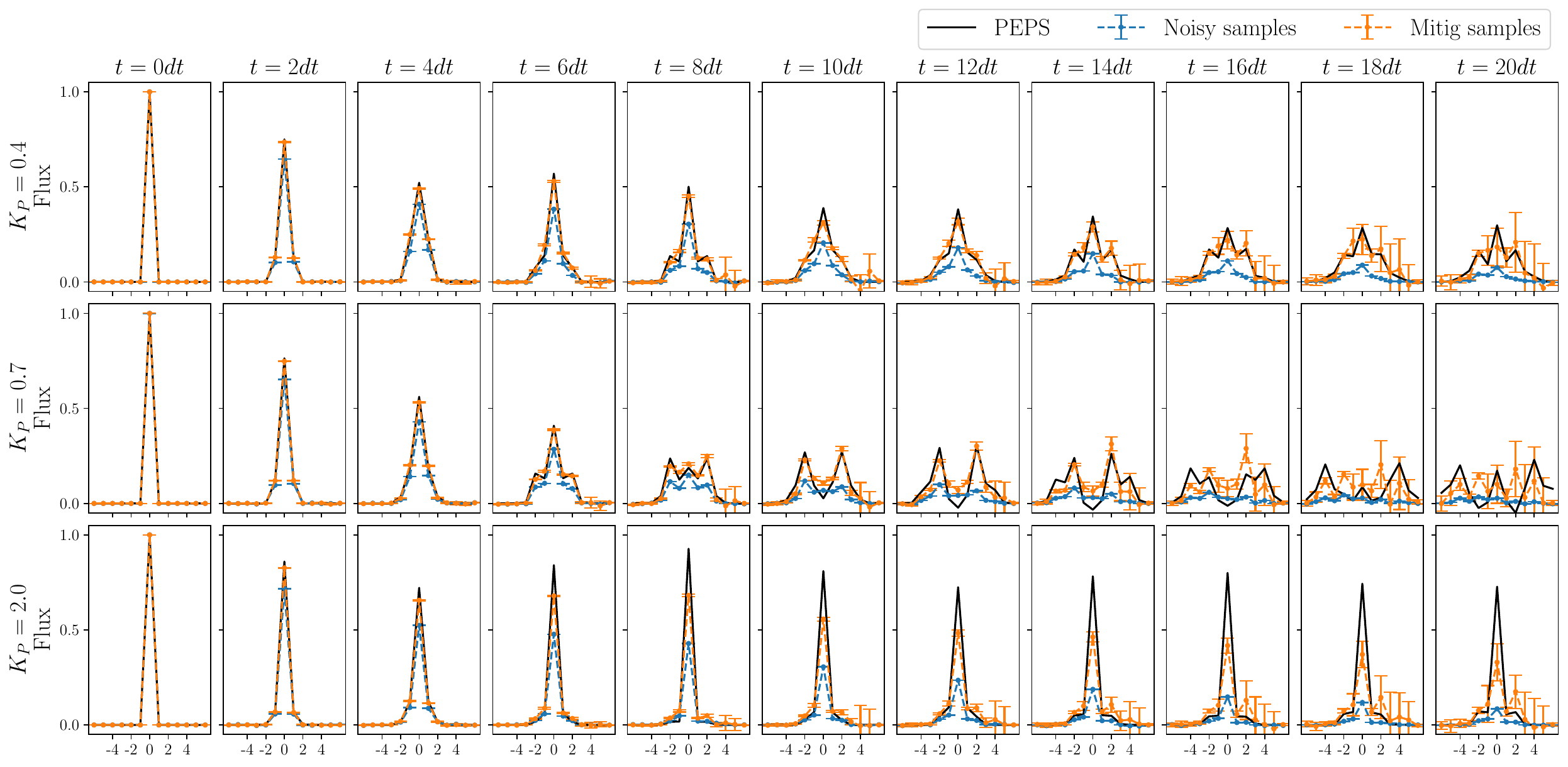}
    \caption{Dynamics of the petal flux density along the cross-section defined in~\Cref{fig:FIG04_flux_imshow} for the larger lattices with 13 full heavy-hexagons, and for the couplings $K_P\in\{0.1, 0.7, 2.0\}$. The observables are measured for various circuit depths corresponding to total times $t=k\times dt$, $k\in\{0,2,\dots,20\}$, and to a time step $dt=0.4$ (in units where the triangle coupling is $1$).}
    \label{fig:FIG11_flux_imshow_bis}
\end{figure}

In this section we provide complementary results for the hardware experiments presented in~\Cref{fig:FIG04_flux_imshow}. We focus on the cross-section depicted in the upper part of~\Cref{fig:FIG04_flux_imshow} and detail the individual fluxes estimated across this slice for the smaller sublattice of 45 qubits, see~\Cref{fig:FIG11_flux_imshow_bis_small}, and for the larger sublattice of 114 qubits, see~\Cref{fig:FIG11_flux_imshow_bis}. We compare both raw expected values, in blue, and mitigated expectation values, in orange, to converged results from tensor network simulations, in black. Both experimental expectation values are represented with statistical errors only, and accounting for the error-mitigation overhead discussed in \Cref{sec:appendix_em_cpt}. Remarkably, we find that the raw expectation values share the qualitative features of the expected noiseless signal, for both subsystem sizes and up to relatively large circuits of up to 16 Trotter layers. The mitigation strategy based on Clifford perturbation theory is able to recover quantitative agreement with the ideal dynamics, although with large error bars for the deeper circuits.

\subsection{Dynamics of height variable order parameters}

\begin{figure}[ht!]
    \centering
    \includegraphics[width=0.6\linewidth]{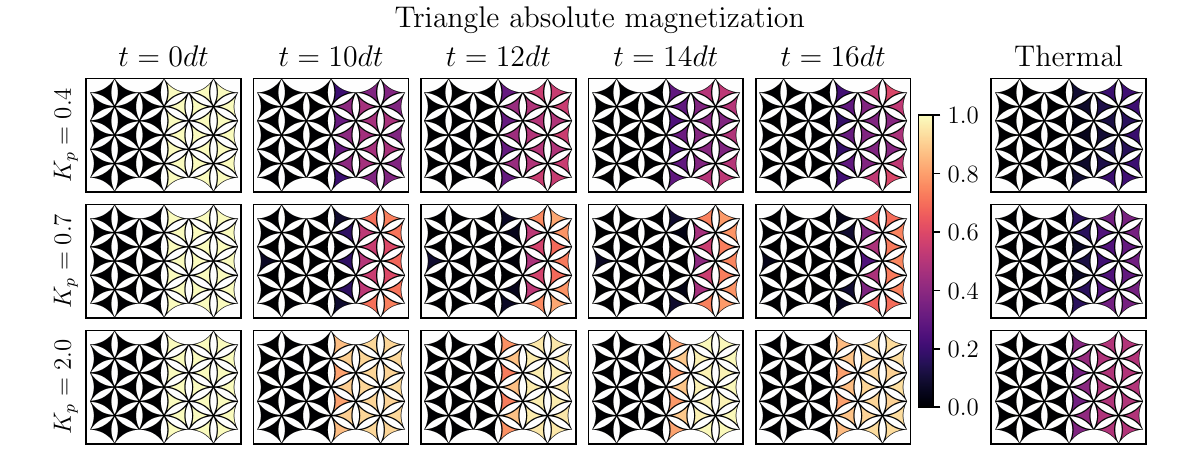}

    \vspace{0.2cm}
    \includegraphics[width=0.6\linewidth]{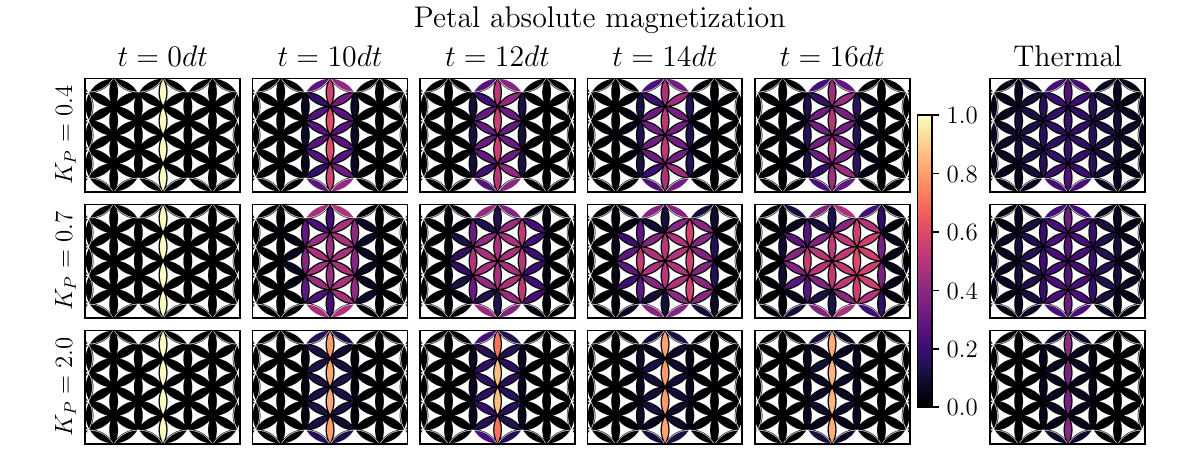}
    \caption{Local ordering of the absolute triangle and petal magnetization $|Z_{t/p}|$ for the three quenching regimes $K_P\in\{0.4, 0.7, 2.0\}$, and for a subset of total times $t=k\times dt$ with $k\in\{0,10,12,14,16\}$. Non-zero eigenvalues for the triangle (resp.\ petal) magnetization correspond to the ordering of the triangle (resp.\ petal) plaquettes. Values at thermal equilibrium (obtained by QMC) are also included. These correspond to the infinite time limit $t=\infty$, assuming thermalization at the quench temperature given by the initial state configuration. }
    \label{fig:FIG13_magnetization}
\end{figure}

In addition to the flippabilities, the same shots from the quantum hardware can be used to evaluate the local ordering of the height variables. This is represented in~\Cref{fig:FIG13_magnetization}, which should be compared to~\Cref{fig:FIG05_flippabilities}. While the flippabilities are agnostic of the redundancy in the height variable representation, the initial state with two static charges corresponds to inserting a boundary wall of the onsite magnetization. In agreement with our discussion from~\Cref{sec:quench}, the quench dynamics at the transition point $K_P=0.7$ are characterized by an ordering of the petals in the interior of the two flux strands.

\section{Quantum Monte Carlo algorithm}\label{sec:appendix_qmc-algorithm}


In this section, we discuss the details of the Quantum Monte Carlo algorithm which was used in this work in order to determine the static properties of the system. The algorithm is quite similar to the cases of the square~\citep{Banerjee201321dU1} and triangular~\citep{Banerjee2022NematicConfinedPhasesa} lattices and follows the general principles discussed, for example, in~\citep{Banerjee2021RecentProgressCluster, Banerjee2018SU2QuantumLink}. For the square lattice, a thorough description of the algorithm can be found in~\citep{Widmer20152+1dU1Quantum}. Since the hexafoil lattice presents certain specific non-trivial difficulties, we highlight relevant details here.

\subsection{Euclidean path integral}
\begin{figure}
\begin{center}
\tdplotsetmaincoords{60}{100}
\begin{tikzpicture}[scale=1.5,tdplot_main_coords]
    \node (Ai) at (0,0,-1) {$A$};
    \node (Af) at (0,0,1) {$A'$};
    \node (B1) at (0,1,0) {$B_1$};
    \node (B2) at ({-sqrt(3)/2},-0.5,0) {$B_2$};
    \node (B3) at ({sqrt(3)/2},-0.5,0) {$B_3$};
    \coordinate (O) at (0,0,0);
    \draw (Ai) -- (Af);
    \draw[dotted] (B1) -- (O);
    \draw[dotted] (B2) -- (O);
    \draw[dotted] (B3) -- (O);
    \draw[dashed] (0,-1,-1) node[left] {\footnotesize Timeslice $2n$} -- (0,-.2,-1);
    \draw[dashed] (0,-1,0) node[left] {\footnotesize Timeslice $2n+1$} -- (0,-.2,0);
    \draw[dashed] (0,-1,1) node[left] {\footnotesize Timeslice $2n+2$} -- (0,-.2,1);
\end{tikzpicture}
\begin{tikzpicture}[scale=1.5,tdplot_main_coords]
    \node (Bi) at (0,0,-1) {$B$};
    \node (Bf) at (0,0,1) {$B'$};
    \node (A1) at ({sqrt(3)/2},-0.5,0) {$A_1$};
    \node (A2) at ({-sqrt(3)/2},0.5,0) {$A_2$};
    \coordinate (O) at (0,0,0);
    \draw (Bi) -- (Bf);
    \draw[dotted] (A1) -- (O);
    \draw[dotted] (A2) -- (O);
    \draw[dashed] (0,-1,-1) node[left] {\footnotesize Timeslice $2n-1$} -- (0,-.2,-1);
    \draw[dashed] (0,-1,0) node[left] {\footnotesize Timeslice $2n$} -- (0,-.2,0);
    \draw[dashed] (0,-1,1) node[left] {\footnotesize Timeslice $2n+1$} -- (0,-.2,1);
\end{tikzpicture}
\end{center}
\caption{The two kinds of interactions in the Euclidean path integral representation used in the Quantum Monte Carlo calculations. Left: interaction I1 corresponding to the matrix element $\left< A | \exp(\delta t \mathbb{P}_t) | A'\right>$. Right: interaction I2 corresponding to the matrix element $\left< B | \exp(\delta t \mathbb{P}_p) | B' \right>$.}
\label{fig:euclidean-interactions}
\end{figure}
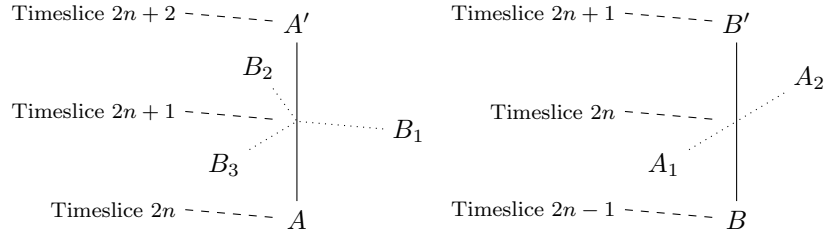

The Quantum Monte Carlo algorithm implements a simulation in Euclidean time $\beta$ directly acting on the dual height variables. The path integral is defined by a Trotter decomposition into $N_T$ imaginary-time steps $\delta t = \beta / N_T$ that act alternately on the triangle and petal variables. As a result, the spacetime lattice has a natural staggered geometry between odd timeslices containing petals and even timeslices containing triangles. As shown in \Cref{fig:euclidean-interactions}, the structure of the Hamiltonian results in two kinds of matrix elements:
\begin{itemize}
    \item \textbf{I1:} The matrix element $\left< A | \exp(-\delta t H_T(t)) | A' \right> = \left< A | \exp(\delta t \mathbb{P}_t X_t) | A'\right>$ between values $A$ and $A'$ of the height variable $h_t$ on triangle $t$ at timeslices $2n$ and $2n+2$, given that the three adjacent petal variables have values $B_1$, $B_2$, and $B_3$,
    \item \textbf{I2:} The matrix element $\left< B | \exp(-\delta t H_P(t)) | B' \right> = \left< B | \exp(\delta t \mathbb{P}_p X_p) | B' \right>$ between values $B$ and $B'$ of the height variable $h_p$ on petal $p$ at timeslices $2n-1$ and $2n+1$, given that the two adjacent triangle variables have values $A_1$ and $A_2$.
\end{itemize}
The path integral weights for the two kinds of interactions are respectively
\begin{align}
    e^{-S_1(t,n)} &= \delta_{B_1, B_2, B_3} \delta_{A, A'} (e^{-\delta t} - 1) + \sinh(\delta t) \delta_{B_1, B_2, B_3} + \delta_{A, A'} \, , \label{eq:S1} \\
    e^{-S_2(p,n)} &= \delta_{A_1, A_2} \delta_{B, B'} (e^{-\delta t K_P} - 1) + \delta_{A_1, A_2} \sinh(\delta t K_P) + \delta_{B, B'} \, , \label{eq:S2}
\end{align}
where $t$ ($p$) indicates the choice of triangle (petal) and $n$ corresponds to the choice of timeslice on which the interaction is localized (see~\Cref{fig:euclidean-interactions}).
Here all the interactions are expressed in terms of combinations of Kronecker deltas applied to the height variables of the sites involved in each interaction. The full path integral representation using a discretized time interval with $N_T$ timeslices is then simply a product over these terms,
\begin{equation}
    Z = \sum_{\{ h_t \}, \{ h_p \}} \left[ \prod_{t \in T} \prod_{n=1}^{N_T} e^{-S_1(t,n)} \right] \left[ \prod_{p \in P} \prod_{n=1}^{N_T} e^{-S_2(p,n)} \right].
\end{equation}

\subsection{Cluster algorithm}
For $K_P \geq 0$, the weights in~\Cref{eq:S1,eq:S2} are positive definite and the path integral can be treated using Monte Carlo sampling. Given the decomposition of the weights into terms involving various products of Kronecker deltas, we are able to apply a cluster algorithm. We refer to previous references for discussions of cluster algorithms in general. Here, we note specific details of our cluster identification and update scheme needed for the heavy-hex lattice on which the height variables are represented.

Because of the structure of the projectors $\mathbb{P}_t$ and $\mathbb{P}_p$ involved in the action terms $S_1$ and $S_2$, respectively, it is not possible to perform a cluster update over all degrees of freedom simultaneously. Instead, we update in ``rounds'' consisting of first sampling bonds, identifying clusters, and applying random cluster flips for only the triangle variables, then performing the same actions for only the petal variables. This is analogous to the rounds of sublattice updates discussed in Refs.~\citep{Banerjee2018SU2QuantumLink,Banerjee2021RecentProgressCluster,Banerjee2022NematicConfinedPhasesa}.

To form clusters, bonds are only ever placed between variables with \emph{consistent} values; except when Dirac strings are included (see below), this means that the variables bonded together must take on the same value. Bonds are placed for the triangle sublattice in two cases, with probabilities derived respectively from~\Cref{eq:S1} and~\Cref{eq:S2}. For temporally-adjacent triangle variables with values $A = A'$ and for spatially-adjacent triangle variables with values $A_1 = A_2$ (see~\Cref{fig:euclidean-interactions}), active bonds are placed with respective probabilities
\begin{equation}
\begin{aligned}
    p(\text{triangle, temporal}) &= \delta_{B_1, B_2, B_3} (1 - \tanh(\delta t)) + (1 - \delta_{B_1, B_2, B_3}), \\
    p(\text{triangle, spatial}) &= \delta_{B, B'} \left( 1 - \frac{1}{\cosh(K_P \delta t)} \right) + (1 - \delta_{B, B'}).
\end{aligned}
\end{equation}
Likewise, bonds are placed for the petal sublattice in two cases, with probabilities also derived respectively from~\Cref{eq:S2} and~\Cref{eq:S1}. For temporally-adjacent petal variables with values $B = B'$ and for spatially-adjacent petal variables with values $B_1 = B_2 = B_3$, active bonds are placed with respective probabilities
\begin{equation}
\begin{aligned}
    p(\text{petal, temporal}) &= \delta_{A_1, A_2} (1 - \tanh(\delta t K_P)) + (1 - \delta_{A_1, A_2}), \\
    p(\text{petal, spatial}) &= \delta_{A, A'}\left( 1 - \frac{1}{\cosh(\delta t)} \right) + (1 - \delta_{A, A'}).
\end{aligned}
\end{equation}
Note that spatial bonds on petals connect and enforce equality between all three spatially adjacent petals. 

Once clusters have been formed for either the triangle or petal sublattice, each cluster can be randomly flipped with independent probability, except when doing so would update variables in the boundaries with fixed values or would cause a violation of Gauss' Law. Clusters that connect to fixed variables in the boundaries are straightforwardly identified. To avoid violations of Gauss' Law, we take the conservative approach of not flipping any clusters that wind at least once around the temporal boundary.

Enforcing Gauss' law in this way preserves the net external charge at each site, as measured by generators $G_v$ of the local \uone~at each primal site $v$. It is therefore possible to insert fixed even charges by choosing appropriate initial states; odd charges cannot directly be inserted in this way because the height variable representation automatically enforces Gauss' law modulo 2, i.e.,~\Cref{eq:gauss_law_native}. Instead, for pairs of odd charges, it is necessary to modify the interactions by inserting a Dirac string between the sites on which the charges are located. The Dirac string modifies the interaction between adjacent triangle and petal height variables separated by the string, such that all Kronecker deltas and consistency conditions use the opposite height variable value whenever such a value is ``brought through'' the Dirac string. Simulations with given choices of Dirac strings are equivalent up to homotopy, as the Dirac string can be locally moved by a change of representation. Some further discussion of Dirac strings for this algorithm was given in~\citep{Widmer20152+1dU1Quantum} in the case of the square lattice.


\subsection{Energy observables}
Other observables we are interested in measuring are the average local energies $E_t = \left< - \mathbb{P}_t X_t \right>$ and $E_p = \left< - \mathbb{P}_p X_p \right>$, as well as the average total energy $E = \left< H \right> = \sum_{t \in T} E_t + \sum_{p \in P} E_p$.
These observables can be defined in the Euclidean path integral by taking a derivative of the path integral with respect to $\beta = N_T \, \delta t$:
\begin{equation}
    \left< H \right> = -\partial_\beta \log{Z}
    = -\frac{1}{Z N_T} \sum_{\{ h_t \}, \{ h_p \}} \frac{\partial}{\partial \delta t} \left[ \prod_{t \in T} \prod_{n=1}^{N_T} e^{-S_1(t,n)} \right] \left[ \prod_{p \in P} \prod_{n=1}^{N_T} e^{-S_2(p,n)} \right].
\end{equation}
Focusing first on one of the triangle terms, the derivative gives
\begin{equation}
    \frac{\partial}{\partial \delta t} e^{-S_1(t,n)} = \delta_{B_1, B_2, B_3} [ \{\tanh(\delta t) \} \cosh(\delta t) \delta_{A, A'} + \{ \coth(\delta t)  \} \sinh(\delta t) (1-\delta_{A,A'})].
\end{equation}
Focusing next on one of the petal terms, the derivative reads
\begin{equation}
\begin{aligned}
    \frac{\partial}{\partial \delta t} e^{-S_2(p,n)} &= \delta_{A_1, A_2} [ \{K_P \tanh(\delta t K_P) \} \cosh(\delta t) \delta_{B B'} + \{ K_P \coth(\delta t K_P) \} \sinh(\delta t)(1-\delta_{BB'})].
\end{aligned}
\end{equation}
In both cases, the results are written in such a way to isolate the additional factor in curly braces that must be accumulated in the Monte Carlo measurements.
Taking these factors together with the corresponding delta functions gives the relevant observable to be measured in a Monte Carlo simulation:
\begin{equation}
\begin{aligned}
    \left< H \right> &= -\frac{1}{N_T} \mathbb{E}\Big[
    \sum_{t} \delta_{B_1, B_2, B_3}[\tanh(\delta t) \delta_{A, A'} + \coth(\delta t) \, (1-\delta_{A, A'})] \\
    &\qquad+ \sum_{p} K_P \delta_{A_1, A_2} [\tanh(\delta t K_P) \delta_{B, B'} + \coth(\delta t K_P) (1-\delta_{B, B'})]
    \Big].
\end{aligned}
\end{equation}
From this, the local Monte Carlo observables can also be defined by evaluating the following on a specific timeslice, or averaging over all timeslices,
\begin{equation}
\begin{aligned}
    E_t &= -\mathbb{E} \Big[ \delta_{B_1, B_2, B_3} [\tanh(\delta t) \delta_{A,A'} + \coth(\delta t)(1 - \delta_{A,A'})] \Big], \\
    \quad
    E_p &= -\mathbb{E} \Big[ K_P \delta_{A_1, A_2} [\tanh(\delta t K_P) \delta_{B, B'} + \coth(\delta t K_P) (1-\delta_{B, B'})] \Big].
\end{aligned}
\end{equation}
Likewise, the local flippabilities are defined as $F_t = \mathbb{E}[\delta_{B_1,B_2,B_3}]$ and $F_p = \mathbb{E}[\delta_{A_1,A_2}]$, which can be evaluated on a specific timeslice or averaged over all timeslices.

\subsection{Additional QMC results for large lattices}
\begin{figure}[ht!]
    \centering
    \includegraphics[width=0.45\linewidth]{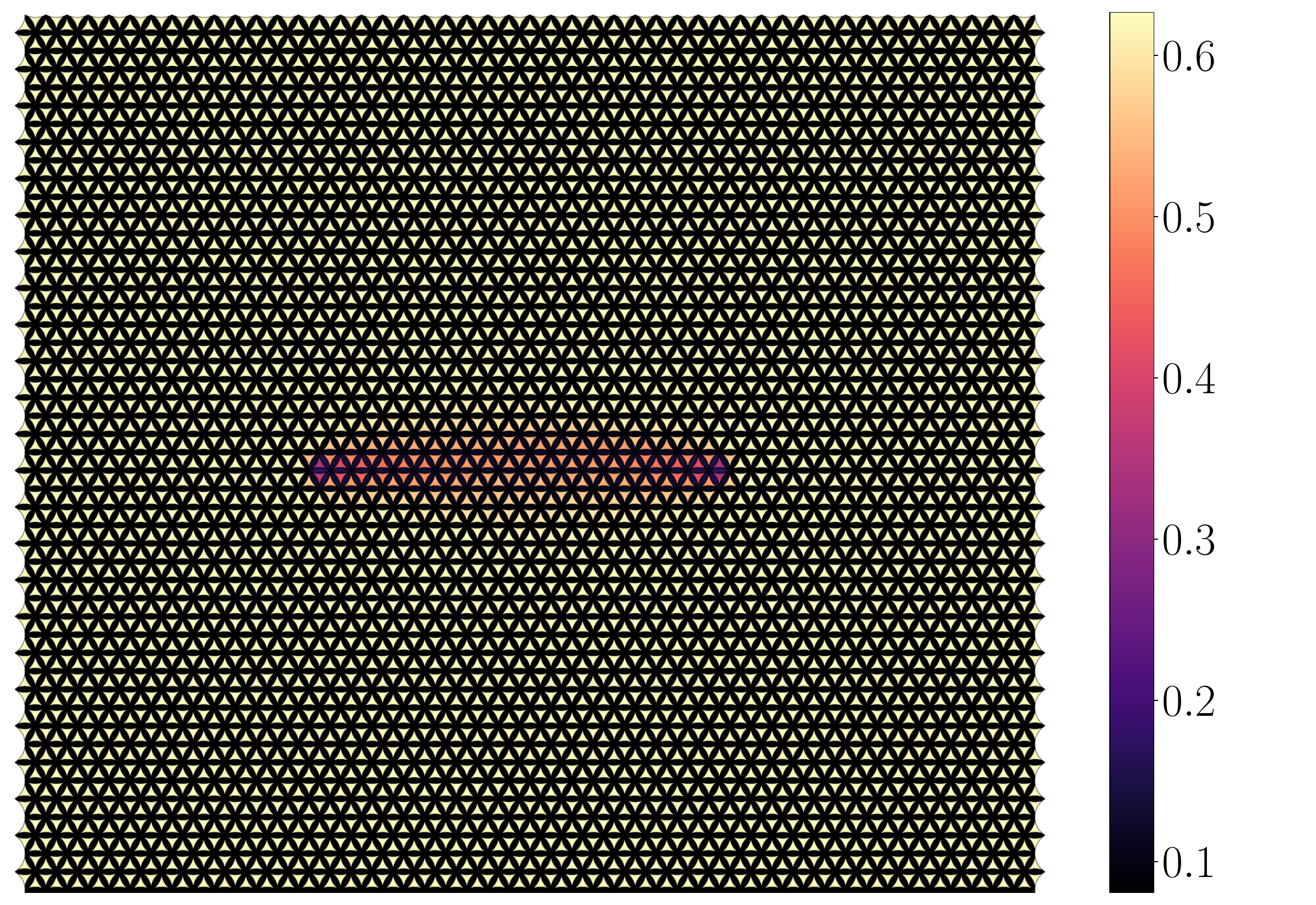}
    \includegraphics[width=0.45\linewidth]{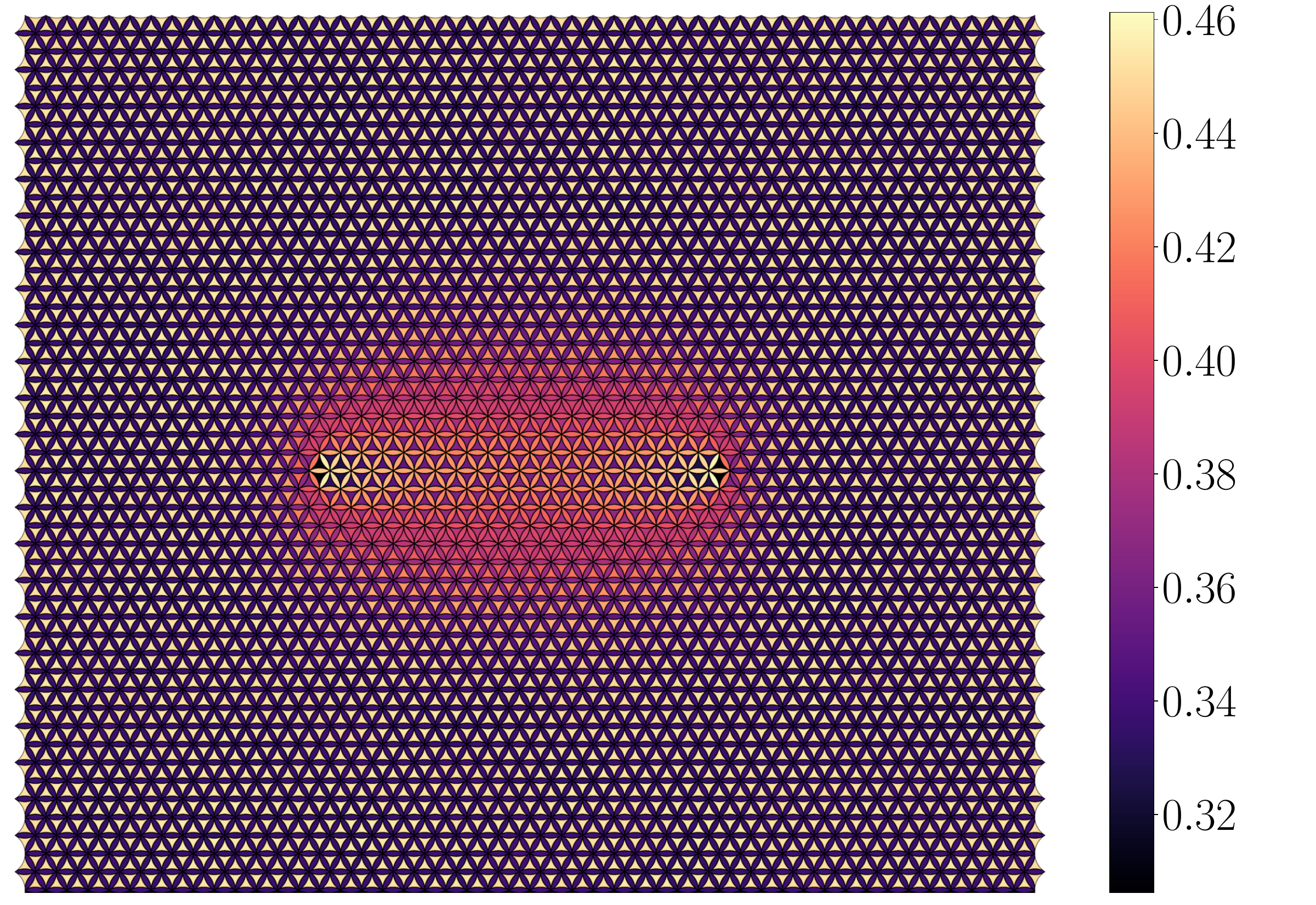}
    \caption{Local energies of petals and triangles for a system of static charges ($q=\pm 1$) at separation $R=20$ and at left) $K_P=0.4$ and right) $K_P=0.7$. The triangle energies are normalized by a factor of $2/3$. Note that the petal energy includes a factor of the coupling, i.e.\ we plot $K_P H_P$.}
    \label{fig:string picture 0.4 energy}
\end{figure}

In addition to the flippabilities presented in~\Cref{fig:string picture 0.4 flippability,fig:string picture 0.7 flippability} in equilibrium for large hexafoil lattices, we represent the corresponding energy densities in~\Cref{fig:string picture 0.4 energy} for completeness. These are computed as part of the Monte Carlo simulation and give a similar picture of the equilibrium ground state in the presence of two external static charges.

\subsection{Additional QMC results for our geometries}
\label{sec:appendix_flux_width}

In addition to the flippabilities presented in~\Cref{fig:FIG05_flippabilities} in equilibrium for our lattice geometries, we represent the corresponding energy densities in~\Cref{fig:string picture 0.4 energy} for completeness. These are computed as part of the Monte Carlo simulation and give a similar picture of the equilibrium ground state in the presence of two external static charges.

\begin{figure*}[ht!]
    \centering
    \includegraphics[width=0.5\textwidth]{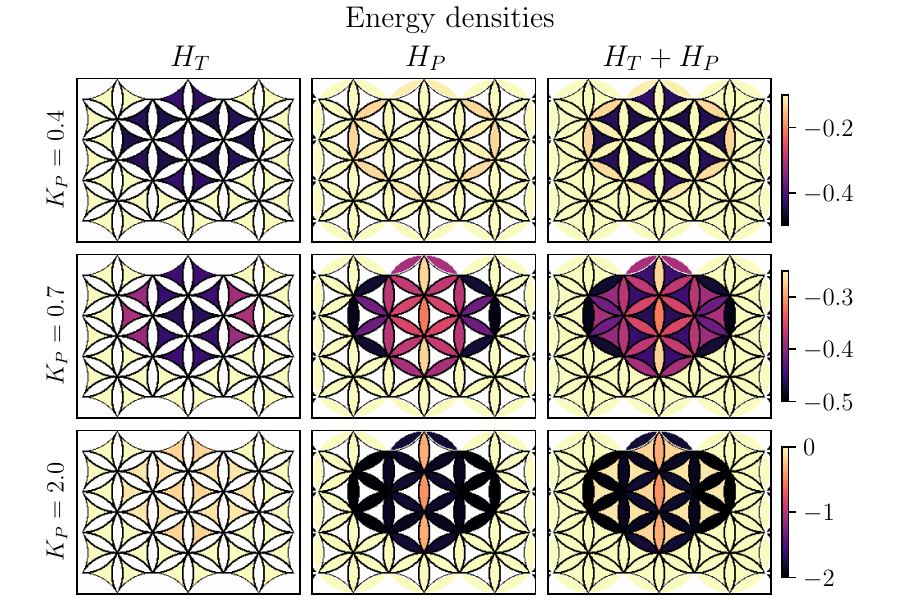}
    \caption{Results for 16T29P10O ensemble across a range of values of $K_P$ on both sides of the transition. The values outside of the sublattice of interest are fixed to 0.0 according to the open boundary conditions.}
\end{figure*}

\begin{figure*}[ht!]
    \centering
    \includegraphics[width=0.5\textwidth]{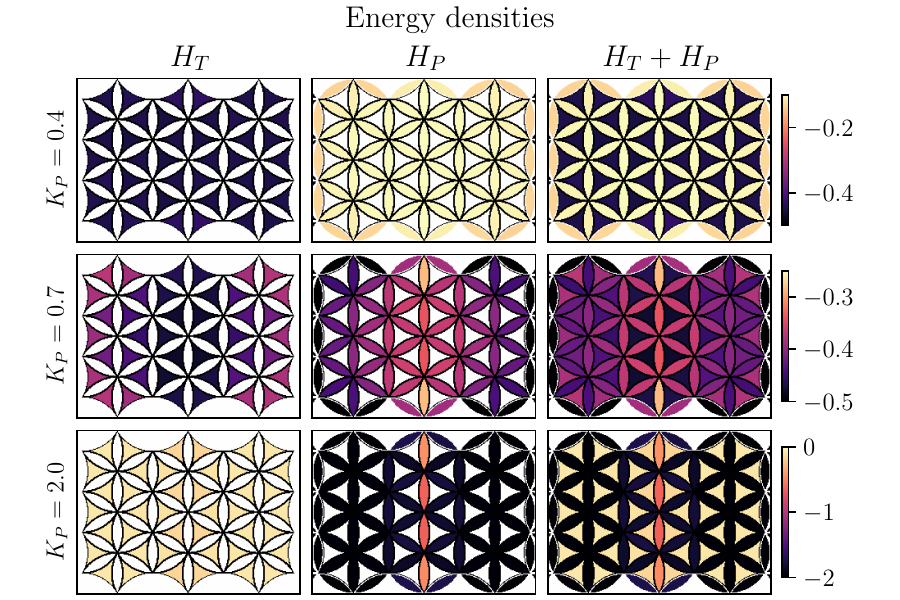}
    \caption{Results for 42T72P18O ensemble across a range of values of $K_P$ on both sides of the transition.}
\end{figure*}

\section{Rokhsar-Kivelson term}\label{sec:rokhsar}

In previous works on the $\mathrm{U}(1)$ quantum link model~\citep{Banerjee201321dU1,Banerjee2022NematicConfinedPhasesa}, a Rokhsar-Kivelson (RK) term~\citep{Rokhsar1988SuperconductivityQuantumHardCorea} was added to the Hamiltonian. In part, this was so that one could study the model as a coupling is varied. In this case, the petals and triangles have different couplings, so this is not needed. In this section, we discuss some explorations of the theory in the presence of a RK term for the petals. This is particularly interesting as it allows one to connect to the triangular lattice model studied in~\citep{Banerjee2022NematicConfinedPhasesa}. As we will see, it does not lead to qualitatively different physics. For the quantum simulations, we have therefore chosen to focus on the model without the RK term, which can be implemented with simpler circuits. The Hamiltonian with the RK term for the petals is given by adding a new gauge-invariant term to the usual Hamiltonian~\Cref{eq:hamiltonian_hexafoil}
\begin{equation}
\begin{aligned}\label{eq:hamiltonian_hexafoil_RK}
    H = - \sum_{t\in T} (U_{t} + U_{t}^\dag) - K_P \sum_{p \in P} (U_{p} + U_{p}^\dag) - 4 K_E \sum_{p=\langle ll'\rangle\in P} E_l E_{l'}  \ ,
\end{aligned}
\end{equation}
where $l, l'$ are the two links making up the petal $p$ and $K_E$ is an arbitrary coupling. The normalization by a factor of four ensures that this term has integer eigenvalues. It can be placed in a more familiar form by noting that 
\begin{equation}
    \label{eq:E E product}
    4 E_l E_{l'} = 2 (U_p+U_p^\dagger)^2-1 \ . 
\end{equation}
Therefore, the terms on either side of~\Cref{eq:E E product} assign $+1$ to flippable petals and $-1$ to non-flippable ones. 

The interesting feature of the RK term for the petals is that it allows one to connect the hexafoil lattice to the triangular lattice. In particular, with our choice of convention for the orientation of the links (i.e.\ ~\Cref{fig:hexafoillattice_dualization}), in the limit $K_E \to +\infty $, the two links on a single petal are forced to agree and every petal is flippable; on the other hand, for $K_E \to -\infty$, the two links on each petal are forced to disagree. Due to our choice of standard orientation, as shown in~\Cref{fig:hexafoillattice_dualization}, in the limit where $K_E\to -\infty$ the theory reduces to that on the triangular lattice studied in~\citep{Banerjee2022NematicConfinedPhasesa}. In principle, one could also add an RK term for the triangles, but this is less interesting. 

\begin{figure}[ht!]
    \centering
    \includegraphics[width=0.3\linewidth]{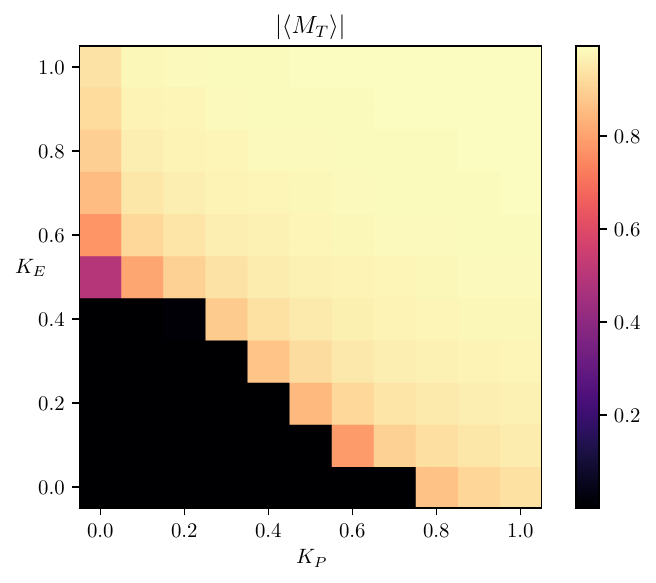}
    \includegraphics[width=0.3\linewidth]{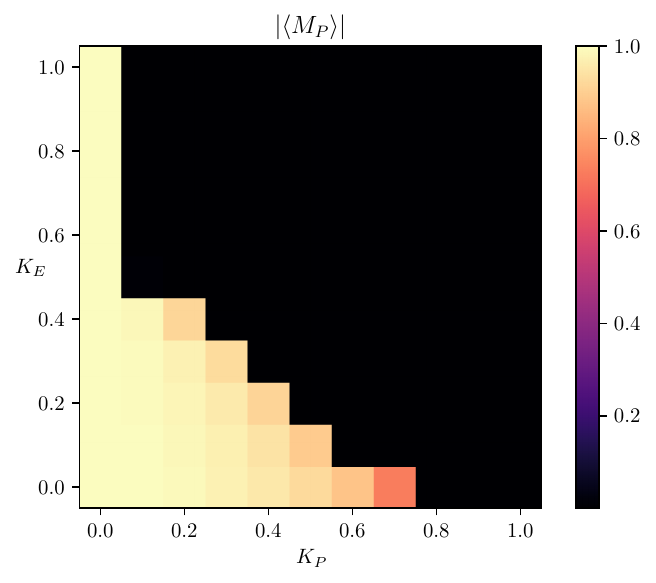}
    \caption{Results of the Quantum Monte Carlo simulations for the magnetizations of triangles and petals, $M_T$ and $M_P$. The two plots are bitmaps showing the values of $\lvert\langle M_T \rangle\lvert$ and $\lvert\langle M_P \rangle\lvert$. For small values of $K_E, K_P$ the petals are ordered while the triangles are disordered. On the other hand, when one of the two couplings becomes large enough, the triangles now order while the petals are disordered.}
    \label{fig:MT MP observables}
\end{figure}

In the dual height description, the Hamiltonian is then given by a modification of the usual Hamiltonian on the heavy-hexagonal lattice~\Cref{eq:hamiltonian_heavyhex} given by the addition of the dual RK term,
\begin{align}\label{eq:hamiltonian_heavyhex_RK}
    H_\mathrm{dual}(K_P, K_E) = - \sum_{t\in T} \mathbb{P}_t S^X_{t} - K_P \sum_{p\in P} \mathbb{P}_p S^X_{p} - K_E \sum_{p\in P} (2 \mathbb{P}_p - 1)\, .
\end{align}
The phase diagram of the theory in the $K_P, K_E$ plane can then be studied via QMC simulations much like for the $K_E=0$ case considered in the main text. To avoid a sign problem, the QMC simulations are effectively limited to the range $K_P, K_E \geq 0$ but can be further extended using the transformations described in~\Cref{sec:model}. Qualitatively, the phase diagram is the same as the one for $K_E=0$ described in~\Cref{fig:MT_MP_history_combined}: one has two different phases where either the triangles or petals order, separated by a first order phase transition. For values of $K_P, K_E$ in the range $[0.0, 1.0]$, the results of QMC simulations on a lattice with $48$ rows and columns of hexagons with periodic boundary conditions are shown in~\Cref{fig:MT MP observables}. The lattice is large enough that tunneling between different vacua is strongly suppressed. For small values of $K_E, K_P$, the petals are ordered while the triangles are disordered; on the other hand, when one of the two couplings becomes large, the opposite happens: the triangles are ordered, while the petals are disordered. In both phases, charge conjugation is spontaneously broken, but no further symmetry distinguishes the two phases. Although less sharp, these basic features of the phase diagram are already visible via exact diagonalization studies on a single hexagon, as discussed in~\Cref{sec:appendix_exact_diagonalization}.

Until now, we have reported results for values of $K_P, K_E$ roughly of order one. These should be compared with the value of the coupling for the triangle plaquettes which is effectively $1$. For large absolute values of $K_P, K_E$ the triangle Hamiltonian can be ignored and the petals fluctuate independently. In this case, the resulting Hamiltonian on each petal can be written as a $4 \times 4$ matrix and diagonalized analytically. This then predicts phase transition lines at $K_E=-\tfrac12 \lvert K_P \lvert$ for $K_E< 0$, with the petals ordered above these lines and disordered below, while the triangles are always disordered. These results are also confirmed by exact diagonalization results on a single hexagon as shown in~\Cref{fig:phase-diagrams-ed} in~\Cref{sec:appendix_exact_diagonalization}, which also summarizes the large $K_P, K_E$ phase diagram. 

\section{Exact diagonalization}\label{sec:appendix_exact_diagonalization}

Preliminary studies of the system were performed via exact diagonalization on a small system of one hexagon with open boundaries, directly using the original link variables on the hexafoil lattice. This system has six triangles and six petals. As remarked in~\Cref{sec:model}, with open boundaries the duality between the link and height variables is exact and therefore exact diagonalization results in the link basis can be used to cross-check results from Quantum Monte Carlo and quantum simulation, which were performed in the dual height basis.

On one hexagon with open boundaries, there are $730$ gauge-invariant states. These were found by brute force computation. As is well-known, even restricting to the gauge-invariant subspace, the number of states grows exponentially with the system size and currently represents the first limitation in scaling exact diagonalization studies. The number of gauge-invariant states on a given lattice can be computed exactly for the formulation of gauge theories for quantum simulation based on finite groups~\citep{Mariani2023HamiltoniansGaugeinvariantHilbert,Mariani2024AlmostGaugeinvariantStates,Mariani2025CountingGaugeinvariantStates}. Unfortunately, for quantum link models, even for just the $s=1/2$ $U(1)$ case, it does not seem to be possible to obtain a simple general formula. This problem is in fact equivalent to an open problem in graph theory~\citep{Beck2006NumberNowherezeroFlows} and closely related to a \texttt{\#P-hard} (i.e.\ intractable) counting problem~\citep{Baldoni-Silva2004CountingIntegerFlows}. On a lattice where each site has the same number of ingoing and outgoing vertices, every configuration of the $s=1/2$ $U(1)$ quantum link model can be mapped to a configuration of the $\mathbb{Z}_2$ lattice gauge theory; since the Gauss law is stricter for quantum link models (i.e.\ it must be satisfied exactly and not modulo two), they have strictly fewer gauge-invariant states, i.e.\ using the formula from~\citep{Mariani2023HamiltoniansGaugeinvariantHilbert}, the dimension of the physical subspace of the quantum link model satisfies
\begin{equation}
    \label{eq:hilbert space bound}
    \mathrm{dim} \mathcal{H}_{\mathrm{phys}} \leq 2^{|E|-|V|+1} \ ,
\end{equation}
where the lattice is assumed to be connected, with $|V|$ sites and $|E|$ links. 

\begin{table}[!ht]
    \centering
    \begin{tabular}{l c c r}
    \multicolumn{1}{c}{Geometry} & Boundaries & \multicolumn{1}{c}{$\mathrm{dim} \mathcal{H}_{\mathrm{phys}}$} \\
    \hline
    1 Hexagons              & open & 730  \\
    2 Hexagons              & open & 865934 \\
    3 Hexagons (triangle)   & open  & 1732986684 \\
    3 Hexagons (line)       & open  & 1027338058 \\
    4 Hexagons (rhombus)    & open  & 3500277676228 \\
    5 Hexagons (trapezoid)  & open  & 7088358951181634
    \end{tabular}
    \caption{Dimension of the gauge-invariant Hilbert space associated with various lattice geometries.}
    \label{tab:dim-H}
\end{table}

In the specific case of the hexafoil lattice, knowing the brute-force value computed from one hexagon, it is possible to extract the dimension of the physical Hilbert space for iteratively larger lattices by imagining attaching an extra hexagon and matching the links on the boundary. The results of this calculation are shown in~\cref{tab:dim-H}. The result for two hexagons was also confirmed via a brute-force search. It should be pointed out that the bound~\Cref{eq:hilbert space bound} is quite loose, pointing to the fact that quantum link models have a much smaller physical Hilbert space than $\mathbb{Z}_N$ gauge theories.

\begin{figure}[t]
    \centering
    \includegraphics[width=0.32\textwidth]{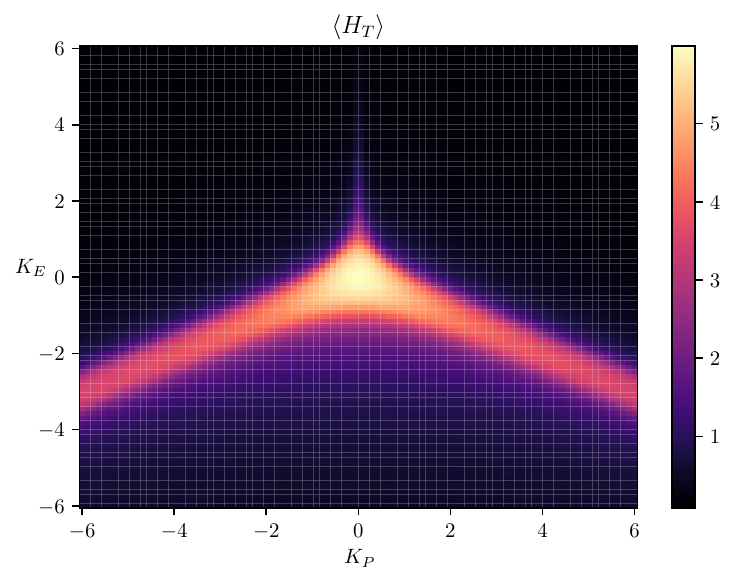}
    \includegraphics[width=0.32\textwidth]{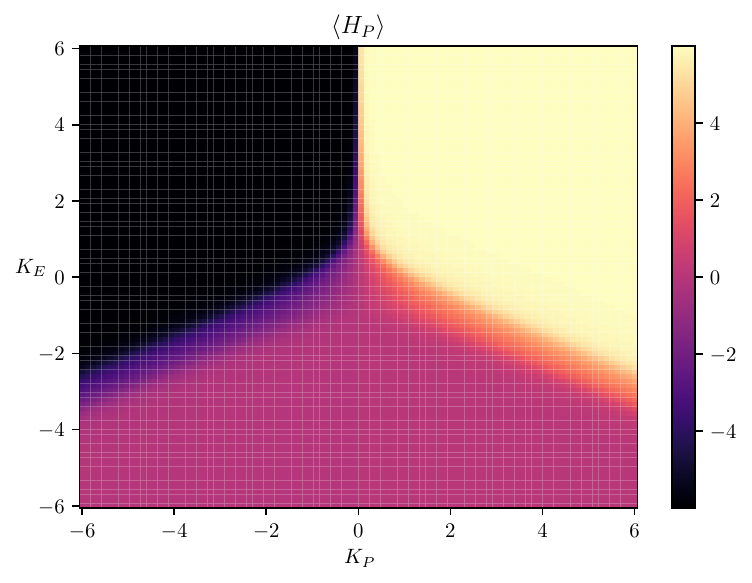}
    \caption{Ground state energy of the triangle and petal Hamiltonians in the $K_P, K_E$ plane, from exact diagonalization on one hexagon with open boundaries.
    }
    \label{fig:phase-diagrams-ed}
\end{figure}

Via exact diagonalization, we studied some features of the phase diagram of the model. It is quite remarkable that, at least at a qualitative level, the main features of the phase diagram are already visible from a single hexagon with open boundaries. This is especially useful, since large scale Quantum Monte Carlo simulations suffer from a sign problem in certain regions of the phase diagram and are therefore effectively limited to $K_E, K_P\geq 0$. In~\Cref{fig:phase-diagrams-ed}, we show the ground state expectation value of the triangle and petal Hamiltonians on one hexagon, covering also the negative $K_E, K_P$ regions. From the figure, one sees that everywhere except near the origin and on two half-lines, the triangles are disordered; on the other hand, the petals order above the two half-lines and are disordered otherwise. This is consistent with the analytical arguments given in~\Cref{sec:rokhsar}, which identify the two half-lines as located at $K_E=-\frac12 \lvert K_P|$, consistent with the figures. In the small $K_E, K_P$ region, one finds a transition between a phase with ordered triangles and one with ordered petals, which is qualitatively the same as the findings from Quantum Monte Carlo in~\Cref{sec:static_phases}; however, on one hexagon the transition is not very sharp and therefore difficult to identify with any precision, owing to the small lattice size.

\section{Dense decomposition of Trotter layers}
\begin{figure}[ht!]
    \centering
    \includegraphics[width=0.5\linewidth]{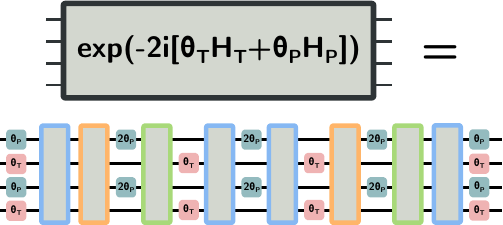}
    \caption{Depth-8 implementation of the combined petal-triangle unitary evolution. The entangling layers (blue, orange, green) as well as the single-qubit $RX$-gates (pink, turquoise) are the same as defined in~\Cref{fig:trotterization}e). The factor of two comes from our attempt to symmetrize the placement of single-qubit rotations in the unitary.}
    \label{fig:FIGA5}
\end{figure}

After having run the circuit depicted in~\Cref{fig:trotterization} on hardware, we discovered a more dense decomposition of the combined Trotter step for both petal and triangle plaquettes. This resulted from the observation that some entangling layers cancel at the boundary between the two unitary evolutions in~\Cref{fig:trotterization}e). We show in~\Cref{fig:FIGA5} a depth-8 implementation of the unitary $\exp(-2i[\theta_P H_P + \theta_T H_T])$. Note that the resulting Trotter error is expected to differ from the decomposition presented in~\Cref{fig:trotterization}e).

\end{document}